\documentclass[12pt,a4paper]{article}
\usepackage{amsmath,amssymb,slashed,cancel}
\usepackage{cite,tabularx}
\usepackage{subcaption}
\usepackage{graphicx,color}
\usepackage[normalem]{ulem}
\usepackage{mathtools}
\usepackage{enumerate}
\allowdisplaybreaks

\usepackage[height=8.85in,width=6.4in]{geometry}
\renewcommand{\baselinestretch}{1.1}
\newcommand\unit[1]{\,\mathrm{#1}}  
\newcommand{\meter}{\,\mathrm{m}}

\newcommand{\GeV}{\,\mathrm{GeV}}

\newcommand\vc[1]{\boldsymbol{#1}}  
\newcommand\w[1]{_{\mathrm{#1}}}  

\newcommand\package[2][\relax]{\texttt{#2\ifx#1\relax\relax\relax\else\,\linebreak[0]#1\fi}}

\newcommand\dd{\mathrm{d}}

\DeclareMathOperator{\ddelta}{\delta}
\DeclareMathOperator{\Acc}{\mathrm{Acc}}

\newcommand\Ebeam{E_{\mathrm{beam}}}

\numberwithin{equation}{section} 
 
\def\beq#1\eeq{\begin{align}#1\end{align}}

\definecolor{BlueViolet}{rgb}{0.2, 0.00, 0.7}
\definecolor{Blue}{rgb}{0.15, 0.00, 0.9}
\usepackage[colorlinks=true,linkcolor=Blue,citecolor=Blue,urlcolor=BlueViolet]{hyperref}

\newcommand\electronpositronfigures[1]{%
  \begin{subfigure}{0.49\textwidth}%
  \captionsetup{margin={25pt,0pt}}%
  \includegraphics[width=\textwidth]{fig/#1_ele.pdf}%
  \subcaption{electron beam dump}\label{fig:#1_ele}%
  \end{subfigure}%
  \begin{subfigure}{0.49\textwidth}%
  \captionsetup{margin={25pt,0pt}}%
  \includegraphics[width=\textwidth]{fig/#1_pos.pdf}%
  \subcaption{positron beam dump}\label{fig:#1_pos}%
  \end{subfigure}%
}

\begin{document}
\begin{titlepage}
\setcounter{page}{0} 

\begin{center}

\hfill {\tt STUPP-21-246}\\

\vskip .55in

\begingroup
\centering
\large\bf New physics searches at\\[.2em]
the ILC positron and electron beam dumps
\endgroup

\vskip .4in

{
  Kento Asai$^{\rm (a,b)}$,
  Sho Iwamoto$^{\rm (c)}$,
  Yasuhito Sakaki$^{\rm (d)}$, and
  Daiki Ueda$^{\rm (a,e)}$
}

\vskip 0.4in

\begingroup\small
\begin{minipage}[t]{0.9\textwidth}
\centering\renewcommand{\arraystretch}{0.9}
\begin{tabular}{c@{\,}l}
$^{\rm(a)}$
& Department of Physics, Faculty of Science, University of Tokyo, \\
& Bunkyo-ku, Tokyo 113--0033, Japan \\[2mm]
$^{\rm (b)}$
& Department of Physics, Faculty of Science, Saitama University, \\
& Sakura-ku, Saitama 338--8570, Japan \\[2mm]
$^{\rm (c)}$
& Institute for Theoretical Physics, ELTE E\"otv\"os Lor\'and University, \\
& P\'azm\'any P\'eter s\'et\'any 1/A, H-1117 Budapest, Hungary \\[2mm]
$^{\rm (d)}$
& High Energy Accelerator Research Organization (KEK),\\
& Tsukuba, Ibaraki 305--0801, Japan \\[2mm]
$^{\rm (e)}$
& Center for High Energy Physics, Peking University, Beijing 100871, China\\
\end{tabular}
\end{minipage}
\endgroup

\end{center}

\vskip .4in

\begin{abstract}\noindent
We study capability of the ILC beam dump experiment to search for new physics, comparing the performance of the electron and positron beam dumps.
The dark photon, axion-like particles, and light scalar bosons are considered as new physics scenarios, where all the important production mechanisms are included: electron-positron pair-annihilation, Primakoff process, and bremsstrahlung productions.

We find that the ILC beam dump experiment has higher sensitivity than past beam dump experiments, with the positron beam dump having slightly better performance for new physics particles which are produced by the electron-positron pair-annihilation.
\end{abstract}
\end{titlepage}

\setcounter{page}{1}
\renewcommand{\thefootnote}{\#\arabic{footnote}}
\setcounter{footnote}{0}

\begingroup
\renewcommand{\baselinestretch}{1} 
\setlength{\parskip}{2pt}          
\hrule
\tableofcontents
\vskip .2in
\hrule
\vskip .4in
\endgroup
\section{Introduction}
\label{sec:introduction}

The International Linear Collider (ILC) experiment~\cite{Behnke:2013xla,Baer:2013cma,Adolphsen:2013jya,Adolphsen:2013kya,Behnke:2013lya} is one of the proposed next-generation electron-positron colliders.
Through observing collisions of high-energy electrons and positrons, we will measure the property of the Standard Model (SM) precisely, such as properties of the Higgs boson, but it also has sensitivity to physics beyond the Standard Model (BSM).

The high-energy electron and positron beams are, after passing the collision point, absorbed into water tanks, called beam dumps.
In the previous works~\cite{Kanemura:2015cxa,Sakaki:2020mqb,Asai:2021xtg}, the possibility to use the electron beam dump as a fixed target experiment is explored.
Searches for hypothetical particles, such as axion-like particles (ALPs), light scalar particles, and leptophilic gauge bosons, were analyzed, and the ILC beam dump experiment was found to be sensitive to such particles, in particular, with small mass and small couplings to SM particles thanks to the large luminosity and high energy of the electron beam.

In this work, we extend and improve the previous works~\cite{Sakaki:2020mqb,Asai:2021xtg}.
Our improvement is threefold.
First, we utilize the positron beam dump as well as the electron beam dump and compare the positron and electron beam dump experiments.
Second, we analyze three BSM models in parallel: dark photons, ALPs, and new light scalars.
As the third improvement, we consider three production processes of the BSM particles: (a) pair-annihilation, (b) Primakoff process, and (c) bremsstrahlung (cf.\ Fig.~\ref{fig:diag}).

\begin{itemize}
 \item[(a)]
Pair-annihilation processes are caused by a positron in the beam or in an electromagnetic shower.
They typically have a larger cross section than the other processes and give a characteristic shape to the parameter space of the new physics that can be explored~\cite{Wojtsekhowski:2009vz,Wojtsekhowski:2012zq,Raggi:2015gza,Alexander:2017rfd,Marsicano:2018krp,Nardi:2018cxi,Marsicano:2018glj,Battaglieri:2021rwp}.
The sensitivity to the new physics depends on whether electron beam dump or positron beam dump is used.
We quantify these differences to study which beam dump to use in the future.

 \item[(b)]
Primakoff processes originate in new physics couplings to photons, which have large intensity in electromagnetic showers.
They tend to 
have high angular acceptance because the angles of the initial photons and generated new particles are very small with respect to the beam axis.
As a result, in our benchmark models, this process plays an important role even though the new physics couplings to photons are loop-induced.

 \item[(c)]
For bremsstrahlung productions, we update previous studies by considering the secondary components of electrons and positrons in the electromagnetic shower, which have not been taken into account in previous studies. This is significant in small coupling regions.
\end{itemize}

This paper is organized as follows.
In Sec.~\ref{sec:beamdump}, we introduce the setup of our proposing experiment at the ILC beam dump and summarize our analysis procedure.
In Sec.~\ref{sec:results}, the three BSM models are introduced and analyzed in each subsections: the dark photon scenario in Sec.~\ref{sec:result-dp},
the ALP scenario in Sec.~\ref{sec:result-dp},
and the light scalar scenario in Sec.~\ref{sec:result-scalar}.
Section~\ref{sec:summary} is devoted to the summary.
In addition, we have several appendices for completeness.
In Appendices~\ref{sec:tracklength} and \ref{sec:xs}, we collect useful formulae for beam dump analyses; Appendix~\ref{sec:tracklength} introduces the track lengths of particles in electromagnetic showers are provided with fitting functions and Appendix~\ref{sec:xs} collects the production cross sections of the BSM particles at beam dump experiments.
In Appendix~\ref{sec:theta-dist}, we check a simplification on the angular acceptance we utilize in Sec.~\ref{sec:results} and compare our pair-annihilation results to those obtained with more exact evaluations.

\begin{figure}
\centering
\includegraphics[width=0.95\textwidth]{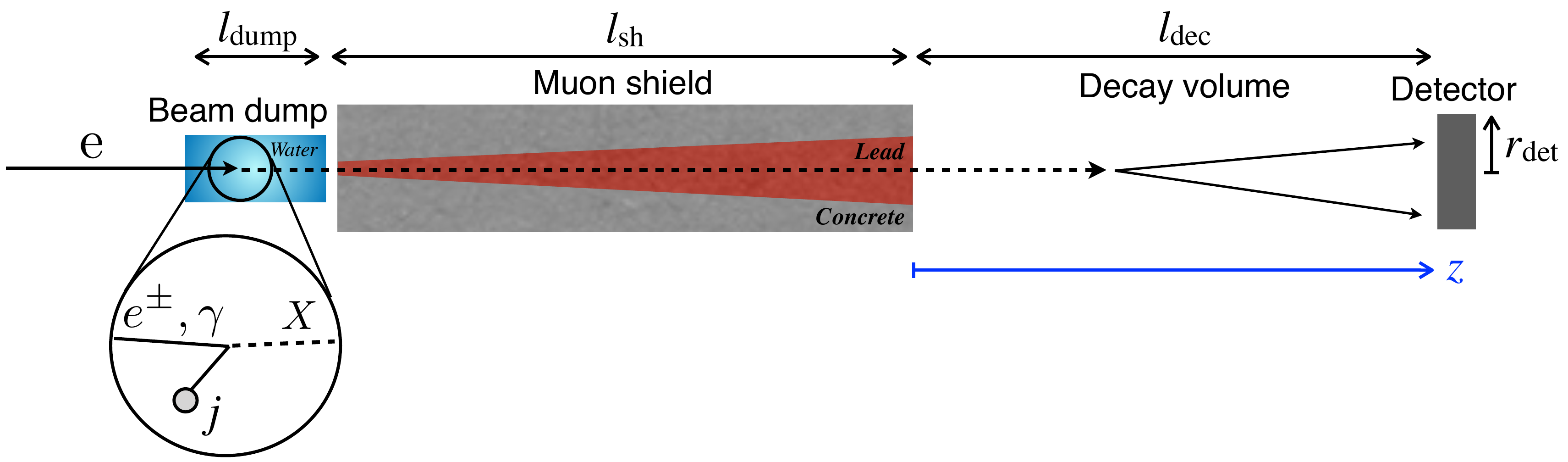}
\caption{A setup for ILC beam dump experiments. It consists of the main beam dump, a muon shield, a decay volume, and a detector.}
\label{fig:exp}
\end{figure}

\section{Beam dump experiment}
\label{sec:beamdump}
The experimental setup is the same as that of Ref.~\cite{Sakaki:2020mqb} and illustrated in Fig.~\ref{fig:exp}.
The ILC main beam dumps are, for both of the electron and positron beams, planned as water cylinders along the beam axes with the length of $l_{\rm dump}=11\meter$~\cite{Satyamurthy:2012zz}.
Our proposal consists of a muon shield with the length of $l_{\rm sh}=70\meter$ made of lead, empty space as the decay volume with $l_{\rm dec}=50\meter$, and a cylindrical detector with the radius of $r_{\rm det}=2\meter$, which are to be installed behind either of the beam dumps.
We consider the 250\,GeV ILC (ILC-250)~\cite{Fujii:2017vwa,Evans:2017rvt} with the beam energy of $\Ebeam=125\GeV$ and
the number of incident electrons and positrons into the beam dump of $N_{e^{\pm}}=4\times 10^{21}/\mathrm{year}$~\cite{Behnke:2013xla,Baer:2013cma,Adolphsen:2013jya,Adolphsen:2013kya,Behnke:2013lya}.
This setup, with its thick shield and very high beam intensity, is particularly sensitive to visible decays of new particles that are weakly coupled to SM particles.

Light BSM particles may be produced in the water beam dump, where the injected beam produces an electromagnetic shower of electrons, positrons, and photons.
They may interact with the water and produce BSM particles.\footnote{%
This interaction may produce muons, which also interact with water in the beam dump or the lead shield to produce BSM particles.
In this work, this muon-originated production of BSM particles is ignored to simplify the analysis.
}
If the BSM particles pass through the muon shield and decay into SM particles in the decay volume, the SM particles may reach the detector and be observed as signal events of our searches.

\begin{figure}
  \centering
  \begin{subfigure}{0.32\textwidth}
  \centering\includegraphics{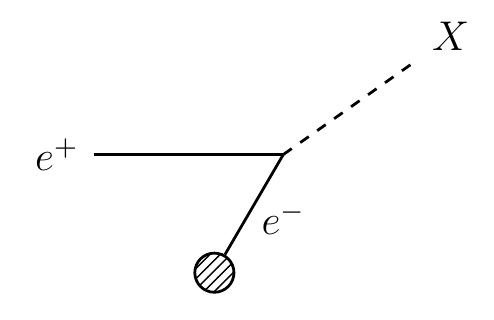}\subcaption{\label{fig:diag-pa}Pair-annihilation}
  \end{subfigure}
  \begin{subfigure}{0.32\textwidth}
  \centering\includegraphics{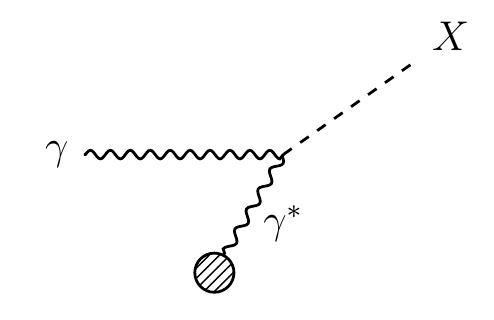}\subcaption{\label{fig:diag-pr}Primakoff process}
  \end{subfigure}
  \begin{subfigure}{0.32\textwidth}
  \centering\includegraphics{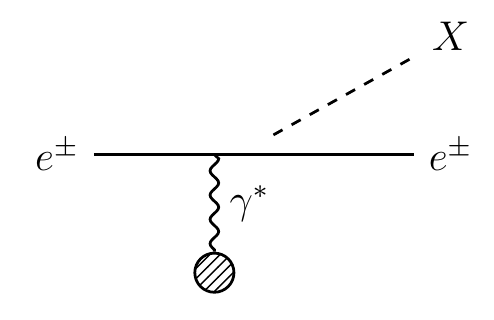}\subcaption{\label{fig:diag-br}Bremsstrahlung}
  \end{subfigure}
\caption{Three production mechanisms of a hypothetical particle $X$.
The left-most particle in each figure comes from the electromagnetic shower or the beam itself, while the blobs illustrate the attached particles originate in water atoms.
}
\label{fig:diag}
\end{figure}

We consider the dark photons, ALPs, and new light scalars as light BSM particles to provide benchmark analyses.
The models are respectively introduced and studied in Secs.~\ref{sec:result-dp}, \ref{sec:result-alp}, and \ref{sec:result-scalar}.
The material-shower interactions to produce BSM particles are, as illustrated in Fig.~\ref{fig:diag}, typically categorized into
bremsstrahlung,
Primakoff process,
and pair annihilation.\footnote{%
We do not discuss here the process $e^+ + e^-\w{atomic} \to X + \gamma$ because it is less important than the other processes when the target is thick. 
The Compton photoproduction, $\gamma\w{shower}+e^-\w{atomic}\to e^-+X$, may also contribute to the production of BSM particles in beam dump experiments (cf.~Ref.\cite{Chakrabarty:2019kdd}).
}

The number of signal events is schematically given by\footnote{We do not consider detector efficiency for simplicity.}
\begin{equation}
    N\w{signal} = N_{e^{\pm}}\times
l_{i}\times n_j\cdot \sigma_{ij\to X}\times \Acc(X)\,,\label{eq:sig}
\end{equation}
where we consider a particle $i$ in the shower interacting with $j$ in the material to produce a BSM particle $X$ (and other SM particles);
the track length $l_i$ of a shower particle $i$ is introduced in Appendix~\ref{sec:tracklength};
the number density of $j$ is denoted by $n_j$, and $\Acc(X)$ is the detector acceptance discussed below.
More specifically,
\begin{subequations}
\begin{align}
 N\w{signal}^{\mathrm{(a)}} &= N_{e^{\pm}}
\int \dd E_{e^+}\,
\frac{\dd l_{e^+}}{\dd E_{e^+}}\cdot n_{e^-}\cdot \sigma(e^+e^-\to X)\cdot 
\Acc(X)\,,
\label{eq:pa}
\\
 N\w{signal}^{\mathrm{(b)}} &= N_{e^{\pm}}
\int \dd E_{\gamma}\,
\frac{\dd l_{\gamma}}{\dd E_{\gamma}}\cdot n_{\mathrm{N}}
\int_0^{\pi}\dd\theta_X\frac{\dd \sigma(\gamma\mathrm{N}\to X\mathrm{N})}{\dd\theta_X}
\cdot \Acc(X)\,,
\label{eq:pr}
\\
 N\w{signal}^{\mathrm{(c)}} &= N_{e^{\pm}}\sum_{i=e^-,e^+}
\int \dd E_{i}\,
\frac{\dd l_{i}}{\dd E_{i}}\cdot n_{\mathrm{N}}
\int\dd E_X \int_0^{\pi}\dd\theta_X
\frac{\dd^2\sigma(i \mathrm{N}\to i X\mathrm{N})}{\dd E_X\,\dd \theta_X}
\cdot \Acc(X)\,
\label{eq:br}
\end{align}
\label{eq:sig-specific}%
\end{subequations}%
for each of the production mechanisms in Fig.~\ref{fig:diag}, i.e., (a) pair-annihilation, (b) Primakoff process, and (c) bremsstrahlung, where N denotes nucleus of the target.\footnote{\label{fn:targetnucleus}%
  The nucleus number density is given by
$n\w N=\rho N\w A/A$ for a target with the density $\rho$ and mass number $A$, where $N\w A=6.02\times10^{23}/\mathrm{g}$.
  Because the relevant cross sections are proportional to $Z^2$ with the atomic number $Z$, we neglect the hydrogen atoms and simply use $Z=8$ and $A=16$ in evaluations.}
The cross sections on the right-hand side are provided in the respective discussion;
$\theta_X$ denotes the emission angle of X with respect to the direction of $i$ in the lab frame.

Precise estimation of the acceptance $\Acc(X)$ requires full Monte Carlo simulation, which we will leave as future works of great interest.
Instead, we estimate it by
\begin{equation}
 \Acc(X) =
\int_0^{l\w{dec}} \dd z\,\frac{\dd P_{\rm dec}}{\dd z}\cdot \Theta \left(r_{\rm det}-r_{\perp}\right)\,,\label{eq:acc}
\end{equation}
where $z$ denotes the decay position of $X$ (see Fig.~\ref{fig:exp}).
We approximate the decay probability of $X$ at $z$ by\footnote{%
  These expression can be obtained by assuming that $X$ is produced at $z=0$ and $\vc p_X$ is almost parallel to the beam axis, where the first assumption does not hold if a muon is the incident particle $i$.}%
\begin{equation}
    \frac{\dd P_{\rm dec}}{\dd z}=\frac{1}{l_X^{\rm (lab)}}\exp\left(-\frac{l_{\rm dump}+l_{\rm sh}+z}{l_X^{\rm (lab)}}\right)\,,\label{eq:decayP}
\end{equation}
where the lab-frame decay length of $X$ is given by
\begin{equation}
    l_X^{\rm (lab)} = \frac{p_X}{m_X} \frac{1}{\Gamma_X}
\end{equation}
with $\Gamma_X$ being the total decay width and $p_X$ the momentum of $X$.
The angular acceptance is taken into account by the Heaviside step function $\Theta$ in Eq.~\eqref{eq:acc}.
We estimate the typical deviation of the SM particles emitted from $X$ from the beam axis as
\begin{align}
    r_{\perp}=\sqrt{\theta_1^2 (l_{\rm dump}+l_{\rm sh}+l_{\rm dec})^2 +\theta_2^2 (l_{\rm dump}+l_{\rm sh}+l_{\rm dec})^2+\theta_3^2 (l_{\rm dec}-z)^2} ~,
\label{eq:rPerp}
\end{align}
where $\theta_1$ is the angle of the particle $i$ with respect to the beam axis,
$\theta_2$ is the production angle of new light particle, i.e., $\theta_2=\theta_X$ (or $0$ for pair-annihilation), and $\theta_3=\pi m_X/(2E_X)$ is the expected decay angle of the SM particles from $X$ with respect to the direction of $X$.
We estimate $\theta_1$ by Monte Carlo simulations (cf.\ Appendix~\ref{sec:tracklength}) and use the mean value
\begin{align}
\theta_1 = \begin{cases}
  16~{\rm mrad}\cdot{\rm GeV}/E_{e^{\pm}} & \text{(for shower electrons and positrons)}, \\
  8~{\rm mrad}\cdot{\rm GeV}/E_{\gamma} & \text{(for shower photons)}
\end{cases}
\label{eq:theta1mean}
\end{align}
in our analyses. This simplification is checked in Appendix~\ref{sec:theta-dist}, where we instead use the full distribution of $\theta_1$ calculated by the Monte Carlo simulation and compare the results for pair-annihilation process.

\section{Examples of detectable new physics}
\label{sec:results}
We evaluate the sensitivity of the ILC electron and positron beam dump experiments to the new light particles using the formula provided in the previous section.
As benchmark models, we consider three models: dark photon, axion-like particles, and light scalar bosons.
The $95\%$ C.L. sensitivities of the ILC beam experiment for these models are shown in the following subsections.
We ignore background events because of the thick shield setup. The 95\% C.L. exclusion then corresponds to $N\w{signal}\ge 3$.

\subsection{Dark photon}\label{sec:result-dp}
The dark photon ${A'}^\mu$ is described by the following Lagrangian:
\begin{align}
\label{eq:Lag-DP}
    \mathcal{L}\supset -\frac{1}{4}F^{(A')}_{\mu\nu}F^{(A')\mu\nu}-\frac{\epsilon}{2}F^{({\rm em})}_{\mu\nu}F^{(A')\mu\nu}+\frac{m_{A'}^2}{2}{A'}_{\mu}{A'}^{\mu}~,
\end{align}
where $F^{({\rm em})}_{\mu\nu}$ and $F^{(A')}_{\mu\nu}$ are field strength tensors of electromagnetic and dark photons, and $m_{A'}$ is the mass of the dark photon.
The second term of Eq.~\eqref{eq:Lag-DP} is the gauge kinetic mixing term between the electromagnetic and dark photons.
Through this mixing, the dark photon couples to the electromagnetic current of the SM particles as
\begin{align}
   \mathcal{L}_{\rm int} \simeq
   - \epsilon e {A'}_\mu j_\mathrm{em}^\mu~,
\end{align}
with $e$ being the electromagnetic charge and $j_\mathrm{em}^\mu$ the electromagnetic current of the SM particles.

Dark photons are produced by pair-annihilation (Fig.~\ref{fig:diag-pa}) and bremsstrahlung (Fig.~\ref{fig:diag-br}), with the production cross sections summarized in Appendix~\ref{sec:xs}.
The partial decay widths of the dark photon are given by
\begin{align}
    \Gamma(A' \to \ell \bar{\ell}) &=
    \frac{1}{3} \alpha \epsilon^2 m_{A'} \left( 1 + \frac{2 m_\ell^2}{m_{A'}^2} \right) \sqrt{1 - \frac{4 m_\ell^2}{m_{A'}^2}}~, \\
    \Gamma(A' \to {\rm hadrons}) &=
    \Gamma(A' \to \mu^+ \mu^-) R(m_{A'}^2)~
\end{align}
with $\alpha$ being the fine structure constant, $m_\ell$ the lepton mass, and $R(s) \equiv \sigma(e^+e^- \to {\rm hadrons}) / \sigma(e^+e^- \to \mu^+\mu^-)$ the ratio between the production cross section of hadronic final states and muon pairs in $e^+e^-$ collisions \cite{10.1093/ptep/ptaa104}.

\begin{figure}[t]
\centering
\electronpositronfigures{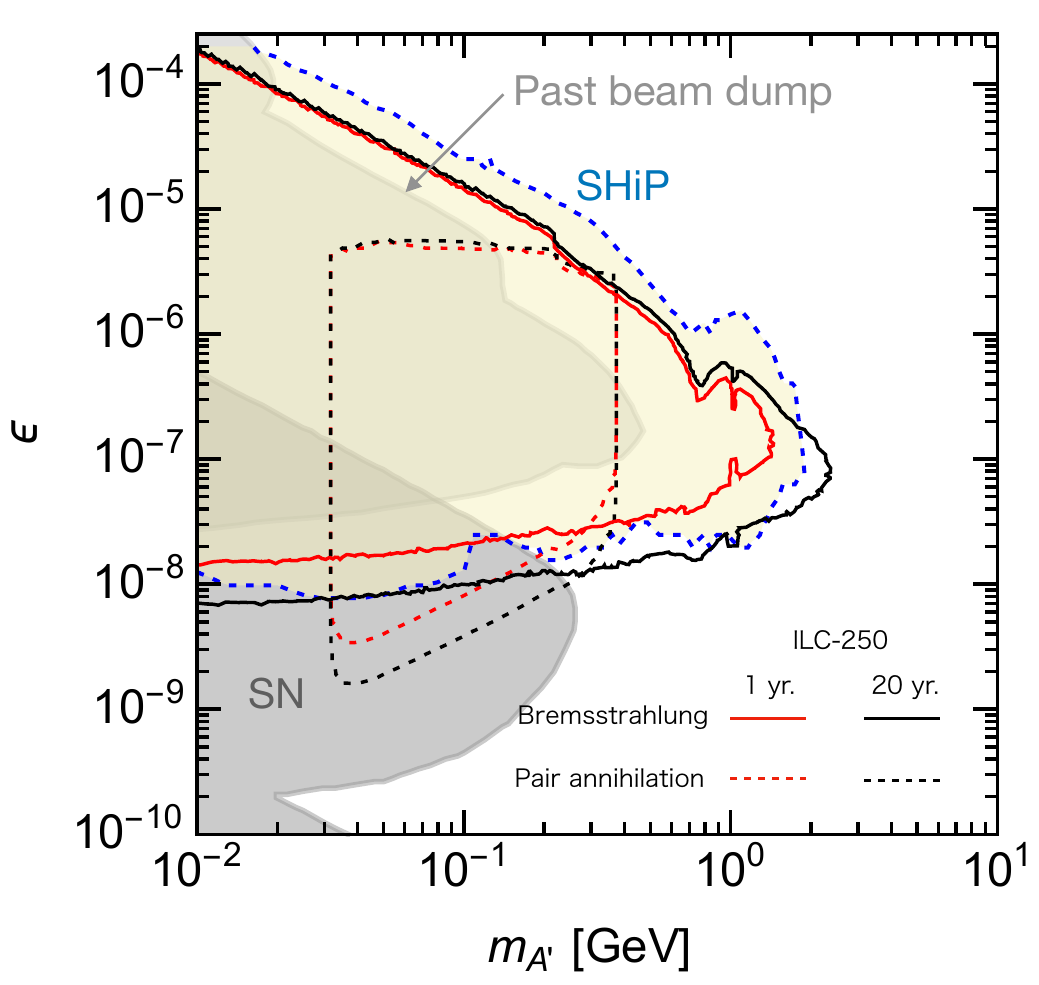}
\caption{
The red and black curves show the bounds of sensitivity for ILC-250 at 95\% C.L. with 1- and 20-year statistics.
The solid lines are the dark photon bremsstrahlung and the dotted lines the dark photon production from the pair annihilation of positrons.
The light-gray shaded region is excluded by past beam dump experiments at 95\% C.L.~\cite{Andreas:2012mt}
The dark-gray shade shows the region excluded by SN~1987A~\cite{Kazanas:2014mca}, while the yellow shade shows the sensitivity of the SHiP experiment~\cite{Anelli:2015pba}.
}
\label{fig:contour_DP}
\end{figure}

Figure~\ref{fig:contour_DP} summarizes the prospects in ILC-250.
The red (black) curves show the expected 95\% C.L. exclusion sensitivity with 1-year (20-year) statistics, based on the respective production mechanism.
The solid lines correspond to the limit obtained by the bremsstrahlung productions, while the dotted lines show the limit from the pair-annihilation process.
The gray-shaded regions are constrained from past beam dump experiments (light gray)\cite{Andreas:2012mt} and supernova bounds (dark gray)\cite{Kazanas:2014mca}.
The yellow-shaded one is the expected sensitivity of SHiP experiment\cite{Anelli:2015pba}.

The bremsstrahlung limit (the solid lines) was previously studied in Ref.~\cite{Kanemura:2015cxa}\footnote{In Ref.~\cite{Kanemura:2015cxa}, the angular acceptance and the positron track length coming from the electromagnetic shower were neglected. }, and it became clear that the ILC beam dump experiment has almost the same sensitivity as the SHiP experiment.
Focusing on small gauge kinetic mixing regions, it can be seen that the pair annihilation of positrons in the ILC beam dump can enlarge the sensitivity of the ILC beam dump experiment\cite{Kanemura:2015cxa} by almost an order of magnitude.

For ease of understanding the dark photon production, we provide approximated formulae of Eq.~\eqref{eq:sig} for three production mechanisms.
\begin{enumerate}[(a)]
\item Pair-annihilation. 
Let us focus on the contour lines in the small $\epsilon$ regions, where the dark photon has a longer lifetime, and the decay probability in Eq.~\eqref{eq:decayP} becomes $\dd P_{\rm dec}/\dd z\simeq 1/l_{A'}^{\rm (lab)}$.
Also, the energy of the produced dark photon is approximately $E^{\rm lab}_{A'}\simeq E_{e^+}\simeq m_{A'}^2/2m_e$.
In the case of the positron beam dump experiment, as shown in Fig.~\ref{fig:track_length} of Appendix~\ref{sec:tracklength}, the primary positron beam contributions to the positron track length becomes comparable to the shower's effects near $ \mathcal{O}(10^{-1})\lesssim E_{e^+}/E_{\rm beam}$, which corresponds to $ \mathcal{O}(10^{-1})~{\rm GeV}\lesssim m_{A'}$.
Then, the number of events is approximately given by
\begin{align}
    N_{\rm signal}\sim \left(\frac{N_{e^{\pm}}}{4\times 10^{21}}\right)\left(\frac{l_{\rm dec}}{50~{\rm m}}\right)
    \left(\frac{\epsilon}{ 10^{-8}}\right)^4\left(\frac{0.2~{\rm GeV}}{m_{A'}}\right)^2,
\end{align}
where we used an approximation: $\dd l_{e^-}/\dd E_{e^-}\simeq (\dd l_{e^-}/\dd E_{e^-})_{\rm primary}\propto  1/E_{e^-}$, $\sigma(e^+ e^-\to A')\propto \epsilon^2$, and $(l^{({\rm lab})}_{A'})^{-1}\propto \epsilon^2 m_{A'}^2/E_{A'}^{{\rm lab}}$.

In the positron beam dump experiment for $m_{A'}\lesssim \mathcal{O}(10^{-1})~{\rm GeV}$, and the electron beam dump experiment, the electromagnetic shower makes dominant in the positron track length.
Then, the number of events is approximately calculated as
\begin{align}
    N_{\rm signal}\sim \left(\frac{N_{e^{\pm}}}{4\times 10^{21}}\right)\left(\frac{E_{\rm beam}}{125~{\rm GeV}}\right)\left(\frac{l_{\rm dec}}{50~{\rm m}}\right)
    \left(\frac{\epsilon}{8\times 10^{-9}}\right)^4 \left(\frac{0.1~{\rm GeV}}{m_{A'}}\right)^4,
\end{align}
where we used approximations: $\dd l_{e^-}/\dd E_{e^-}\simeq (\dd l_{e^-}/\dd E_{e^-})_{\rm shower}\propto  E_{\rm beam}/E_{e^-}^{2}$, $\sigma(e^+ +e^-\to A')\propto \epsilon^2$, and $(l_{A'}^{({\rm lab})})^{-1}\propto \epsilon^2 m_{A'}^2/E_{A'}^{\rm lab}$.

Because of the angular acceptance in Eq.~\eqref{eq:acc}, the positron energy in the beam dump less than $(16~{\rm MeV})\times (l_{\rm dump}+l_{\rm sh}+l_{\rm dec})/r_{\rm det}$ is not detectable, and the dark photon mass less than $\sqrt{2 m_e E_{e^+,{\rm min}}}$ is excluded. %
Also, in the larger coupling regions,  where $l_{A'}^{\rm (lab) }\ll l_{\rm sh}$, the shape of the upper side of dotted contour lines in Fig.~\ref{fig:contour_DP} is determined by the exponential factor in Eq.~\eqref{eq:decayP}.
Then, the contour lines are characterized by
\begin{align}
    \frac{m_{A'} \Gamma_{A'}}{E_{A'}^{\rm lab}}(l_{\rm dump}+l_{\rm sh})\sim {\rm const}.\label{eq:uppline}
\end{align}
Combing $E_{A'}^{\rm lab}\simeq m_{A'}^2/2m_e$ and $\Gamma_{A'}\propto \epsilon^2 \cdot m_{A'}$, Eq.~\eqref{eq:uppline} becomes $\epsilon^2 \sim {\rm const}$.
This means that the upper side of the dotted contour lines does not depend on the mass $m_{A'}$.
\setcounter{enumi}{2}
\item Bremsstrahlung. 
The number of events in the small $\epsilon$ regions is estimated as
\begin{equation}
    N_{\rm signal}\sim \left(\frac{N_{e^{\pm}}}{4\times 10^{21}}\right)\left(\frac{E_{\rm beam}}{125~{\rm GeV}}\right)\left(\frac{l_{\rm dec}}{50~{\rm m}}\right)\left(\frac{r_{\rm det}}{2~{\rm m}}\right)\left(\frac{131~{\rm m}}{l_{\rm dump}+l_{\rm sh}+l_{\rm dec}}\right)\left(\frac{\epsilon}{ 10^{-8}}\right)^4
\end{equation}
with the following approximations: $\dd l_{e^{\pm}}/\dd E_{e^{\pm}}\simeq (\dd l_{e^{\pm}}/\dd E_{e^{\pm}})_{\rm shower}\propto E_{\rm beam}/E_{e^-}^2$, $\sigma (e^{\pm}\mathrm{N}\to e^{\pm}A'\mathrm{N})\propto \epsilon^2 /m_{A'}^{2}$, and $(l_{A'}^{({\rm lab})})^{-1}\propto \epsilon^2 m_X^2/E_{A'}^{\rm lab}$.
Because of the cancellation of $m_{A'}$ between the cross section and the decay length, the number of signals does not much depend on $m_{A'}$.  

Similar to the pair annihilation process, the contour lines in the larger coupling regions behave as Eq.~\eqref{eq:uppline}.
In contrast to the pair annihilation process, $E_{A'}^{\rm lab}$ is not proportional to $m_{A'}$, and Eq.~\eqref{eq:uppline} becomes $\epsilon^2 m_{A'}^2\sim {\rm const}$.

\end{enumerate}

\subsection{ALPs}\label{sec:result-alp}
The Lagrangian related to the ALP $a$ is written as follows~:
\begin{align}
\label{eq:Lag-ALP}
    \mathcal{L} &\supset
    \frac{1}{2} \partial_\mu a \partial^\mu a-\frac{1}{2}m_a^2 a^2 + \sum_{\ell=e,\mu,\tau}\frac{1}{2} c_{a\ell \ell} \frac{\partial_\mu a}{\Lambda} \bar{\ell} \gamma^\mu \gamma_5 \ell -\frac{1}{4}g_{a\gamma\gamma} a F_{\mu\nu}\tilde{F}^{\mu\nu}
    ~
\end{align}
with $m_a$ being the ALP mass, $c_{a\ell \ell}$ the coupling to the SM charged leptons, $\Lambda$ the characteristic breaking scale of the global U(1) symmetry, and $g_{a\gamma\gamma}$ the axion-photon coupling constant.
Here and hereafter, we assume that the coupling of the ALP to the photons arises by the loop corrections from the charged SM leptons. 
Then, the axion-photon coupling constant is obtained as\cite{Bauer:2017ris}
\begin{align}
\label{eq:Lag-ALP-agg}
g_{a\gamma\gamma}=-\frac{\alpha}{\pi}\sum_{\ell=e,\mu,\tau}  \frac{c_{a\ell\ell}}{\Lambda}\cdot B_1(x_{\ell}),
\end{align}
where $x_{\ell}\equiv 4 m_{\ell}^2/m_a^2$, and the loop function is $B_1(\tau)=1-\tau\cdot f(\tau)^2$ with
\begin{align}
\label{eq:loop-func}
f(\tau)=
    \begin{cases}
    {\rm arcsin}\frac{1}{\sqrt{\tau}} &(\tau\geq 1),\\
    \frac{\pi}{2}+\frac{i}{2}\ln \frac{1+\sqrt{1-\tau}}{1-\sqrt{1-\tau}} &(\tau<1).
    \end{cases}
\end{align}
The loop function behaves as $B_1(4m_{\ell}^2/m_a^2)\simeq 1$ for $m_{\ell}\ll m_a$, while $B_1(4m_{\ell}^2/m_a^2)\simeq -m_a^2/12 m_{\ell}^2$ for $m_a\ll m_{\ell}$.

\begin{figure}[p]
\centering
\electronpositronfigures{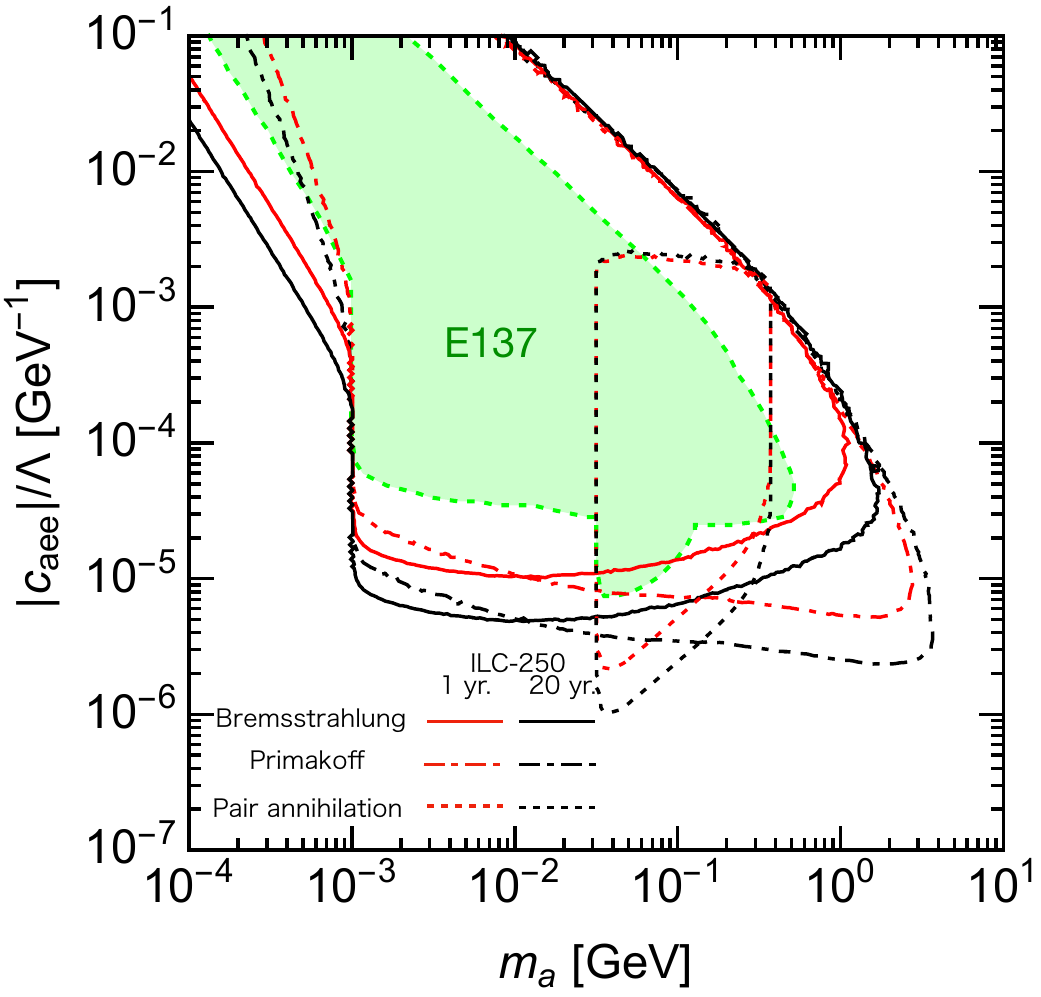}
\caption{%
The result of Case I.
The red and black curves show the bounds of sensitivity for ILC-250 at 95\% C.L. with 1- and 20-year statistics.
The solid lines are the ALP bremsstrahlung production, the dash-dotted lines are the Primakoff process, and the dotted lines the ALP production from the pair annihilation of positrons.
The shaded regions are constraints from the E137 experiments.  
}
\label{fig:contour_ALP}
\end{figure}

\begin{figure}[p]
\centering
\electronpositronfigures{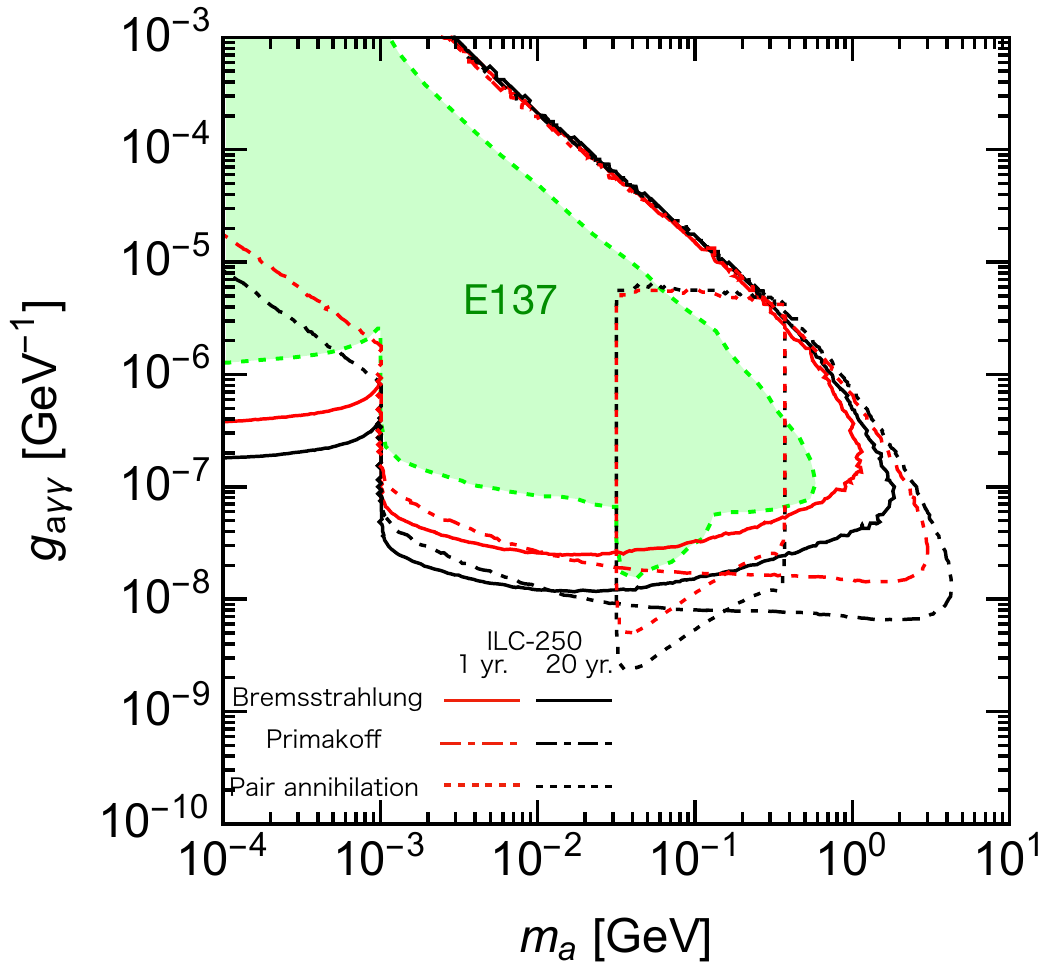}
\caption{%
The same plot as Fig.~\ref{fig:contour_ALP} but in the $(m_a, g_{a\gamma\gamma})$ plane.
}
\label{fig:contour_ALPscale}
\end{figure}

\begin{figure}[p]
\centering
\electronpositronfigures{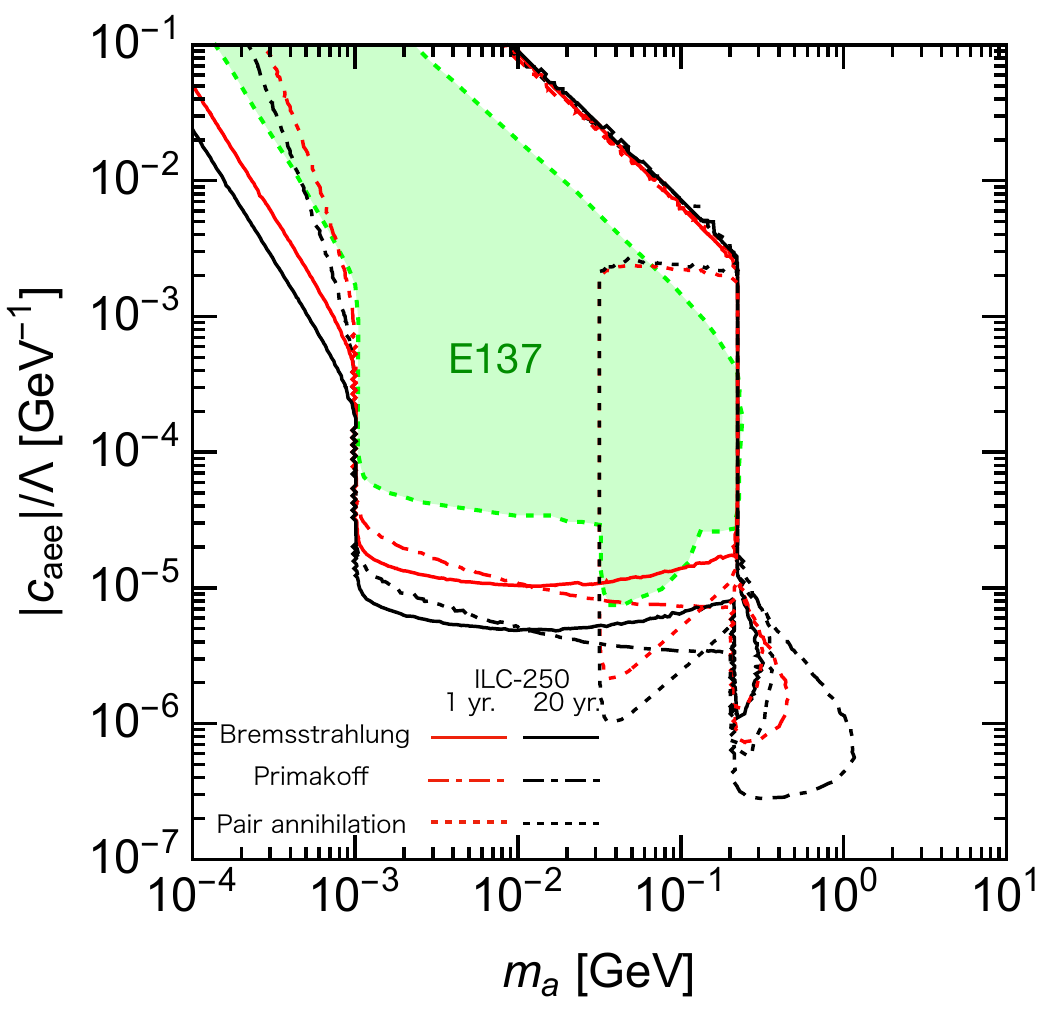}
\caption{%
The same plot as Fig.~\ref{fig:contour_ALP} but for Case II.
}
\label{fig:contour_ALP2}
\end{figure}

\begin{figure}[p]
\centering
\electronpositronfigures{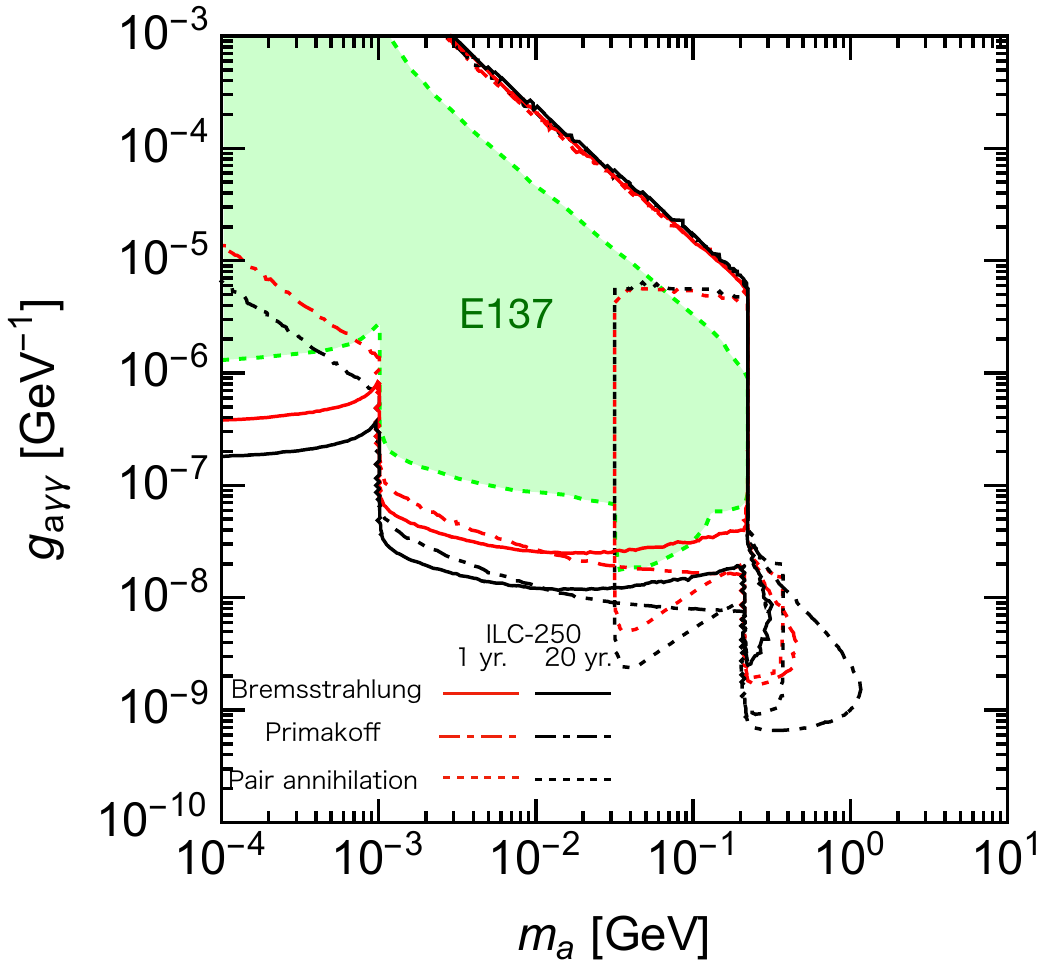}
\caption{%
The same plot as Fig.~\ref{fig:contour_ALP2} but in the $(m_a, g_{a\gamma\gamma})$ plane.
}
\label{fig:contour_ALPscale2}
\end{figure}

ALPs are produced by all the three production mechanisms in Fig.~\ref{fig:diag}; the production cross sections are summarized in Appendix~\ref{sec:xs}.
The decay width of the ALP is obtained as
\begin{align}
    \Gamma(a\to\ell\bar{\ell}) &=
    \frac{m_a m_{\ell}^2}{8 \pi \Lambda^2} |c_{a\ell\ell}|^2 \sqrt{1 - \frac{4 m_{\ell}^2}{m_a^2}}~,
    \\
    \Gamma(a\to\gamma\gamma)&=\frac{\left|g_{a\gamma\gamma}\right|^2 m_a^3}{64\pi}.
\end{align}
As a benchmark, we consider two cases:
\begin{equation}
\begin{aligned}
&c_{aee}\neq 0, ~
 c_{a\mu\mu}=c_{a\tau\tau}=0
&& \text{(Case I),}\\
&c_{aee}=c_{a\mu\mu}=c_{a\tau\tau}
&& \text{(Case II).}
\end{aligned}
\end{equation}

The simulated result for Case I is shown in Fig.~\ref{fig:contour_ALP} with a similar notation as in the dark photon case (Fig.~\ref{fig:contour_DP}).
The solid, dotted, and dot-dashed lines are respectively from the bremsstrahlung, pair annihilation of positrons, and Primakoff processes.
The green shaded region shows the 95\% C.L.\ exclusion by the E137 experiment~\cite{Bjorken:1988as}, which we calculated under their setup with considering the pair-annihilation, Primakoff, and bremsstrahlung processes.\footnote{%
We checked that our results are consistent with those in Ref.~\cite{Dolan:2017osp}.}
Figure~\ref{fig:contour_ALP2} is the same plot as Fig.~\ref{fig:contour_ALP} but for Case II.
In $m_a <2 m_{\mu}$, the contour lines are almost the same as Case I.
While, in $2 m_{\mu}\leq m_a$, the decay mode into a muon pair opens, and the decay length of the ALP becomes shorter.
Then, the probability of decaying particle passing through the muon shield increase if the coupling is small.
Consequently, a constraint region in $2 m_{\mu}\leq m_a$ appears.
 For ease of comparison with other studies, the simulated results in the $(m_a,g_{a\gamma\gamma})$ plane are shown in Figs.~\ref{fig:contour_ALPscale} and \ref{fig:contour_ALPscale2}, which are translated from the $(m_a,|c_{aee}|/\Lambda)$ plane by using Eq.~\eqref{eq:Lag-ALP-agg} and intrinsically the same plots as Figs.~\ref{fig:contour_ALP} and \ref{fig:contour_ALP2}, respectively.
For both cases, it can be seen that the ILC beam dump experiment has higher sensitivity than the E137 experiment in all parameter region because of a large amount of shower photons.
These sensitivities are highly complimentary to the future lepton collider experiments\cite{Bauer:2018uxu,Mimasu:2014nea,Steinberg:2021iay}. The ALPs are produced at $e^+e^-$ colliders by processes $e^+ e^-\to a\gamma, aZ, aH$, and larger coupling and mass regions compared with the beam dump experiment will be constrained.

To understand the parameter dependence of the contour lines in Fig.~\ref{fig:contour_ALP}, we provide approximated formulae of Eq.~\eqref{eq:sig} for three production mechanisms.
\begin{enumerate}[(a)]
\item Pair-annihilation. With the narrow-width approximation, the energy of the produced ALP is $E_a^{\rm lab}\simeq E_{e^+}\simeq m_a^2/2 m_e$.
According to the acceptance in Eq.~\eqref{eq:acc}, the incident positron energy less than $\left(16~{\rm MeV}\right)\times (l_{\rm dump}+l_{\rm sh}+l_{\rm dec})/r_{\rm det}$ is not detectable, and the ALP with the mass less than $ \sqrt{2 m_e E_{e^+,{\rm min}}}$ is excluded.

Let us consider the smaller coupling regions in the positron beam dump experiment.
According to Fig.~\ref{fig:track_length} in Appendix.~\ref{sec:tracklength}, the primary positron track length becomes comparable to the electromagnetic shower's effects for $\mathcal{O}(10^{-1})\lesssim E_{e^+}/E_{\rm beam}$ corresponding to $\mathcal{O}(10^{-1})~{\rm GeV}\lesssim m_a$. 
Then, the number of events in the small coupling regions is approximately obtained as
\begin{align}
    N_{\rm signal}\sim \left(\frac{N_{e^{\pm}}}{4\times 10^{21}}\right)\left(\frac{l_{\rm dec}}{50~{\rm m}}\right)
    \left(\frac{\left|c_{aee}/\Lambda\right|}{10^{-5}~{\rm GeV}^{-1}}\right)^4 \left(\frac{0.2~{\rm GeV}}{m_a}\right)^{2},
\end{align}
where $\dd P_{\rm dec}/\dd z\simeq 1/l^{({\rm lab})}_a$, $\dd l_{e^-}/\dd E_{e^-} \simeq (\dd l_{e^-}/\dd E_{e^-})_{\rm primary}\propto 1/E_{e^-}$, $\sigma (e^+ e^-\to a)\propto\left|c_{aee}/\Lambda\right|^2$, and $(l^{({\rm lab})}_a)^{-1}\propto \left|c_{aee}/\Lambda\right|^2 m_a^2/E_a^{\rm lab}$ are used.

In the positron beam dump experiment for $m_a\leq \mathcal{O}(10^{-1})~{\rm GeV}$, and the electron beam dump experiment, the electromagnetic shower is dominant in the positron track length, and the number of events in the small coupling regions is approximately calculated as
\begin{align}
    N_{\rm signal}\sim \left(\frac{N_{e^{\pm}}}{4\times 10^{21}}\right)\left(\frac{E_{\rm beam}}{125~{\rm GeV}}\right)\left(\frac{l_{\rm dec}}{50~{\rm m}}\right)\left(\frac{\left|c_{aee} /\Lambda\right|}{5\times 10^{-6}~{\rm GeV}^{-1}}\right)^4 \left(\frac{0.1~{\rm GeV}}{m_a}\right)^{4},
\end{align}
where $\dd P_{\rm dec}/\dd z\simeq 1/l^{({\rm lab})}_a$, $\dd l_{e^-}/\dd E_{e^-} \simeq (\dd l_{e^-}/\dd E_{e^-})_{\rm shower}\propto E_{\rm beam}/E^{2}_{e^-}$, $\sigma (e^+ e^-\to a)\propto\left|c_{aee}/\Lambda\right|^2$, and $(l^{({\rm lab})}_a)^{-1}\propto \left|c_{aee}/\Lambda\right|^2 m_a^2/E_a^{\rm lab}$ are used.

In the Case II, the ALP decay into a muon pair for $2 m_{\mu}\leq m_a$, and the decay length of the ALP becomes shorter.
Consequently, a constraint region in smaller coupling arises.

In the larger coupling regions, because of the acceptance in Eq.~\eqref{eq:acc}, the contour lines are characterized by
\begin{align}
    \frac{m_a \Gamma_a}{E_a^{\rm lab}}(l_{\rm dump}+l_{\rm sh})\sim{\rm const}.\label{eq:upplineALP}
\end{align}
For $m_a < 1.2\times 10^3 m_e$, the decay mode into a electron pair makes dominant in the total decay width, and $\Gamma_a\simeq \Gamma(a\to e^+ e^-)$.
Using $E_a^{\rm lab}\simeq m_a^2/2 m_e$, and $\Gamma_a\propto \left|c_{all}\right|^2 m_a$ for $m_a < 1.2\times 10^3 m_e$, Eq.~\eqref{eq:upplineALP} becomes $\left|c_{all}\right|^2\sim {\rm const}$, and the upper contour lines in Fig.~\ref{fig:contour_ALP} and \ref{fig:contour_ALP2} do not depend on $m_a$.

\item Primakoff process. In the small coupling regions, the decay probability in Eq.~\eqref{eq:decayP} becomes $\dd P_{\rm dec}/\dd z\simeq (l^{({\rm lab})}_X)^{-1}$. 
For $m_a<2 m_e$, the ALP only decays to two photons, and $\theta_1$ is greater than $\theta_2$ and $\theta_3$. As calculated in Ref.~\cite{Sakaki:2020mqb}, the minimum value of incident photon energy becomes $E_{i,{\rm min}}=8~{\rm MeV}\times (l_{\rm dump}+l_{\rm sh}+l_{\rm dec})/r_{\rm det}$ because of the acceptance in Eq.~\eqref{eq:acc}. 
Then, the number of signals is approximately calculated as
\begin{align}
    N_{\rm signal}&\sim \left(\frac{E_{\rm beam}}{125~{\rm GeV}}\right)\left(\frac{N_{e^{\pm}}}{4\times 10^{21}}\right)\left(\frac{l_{\rm dec}}{50~{\rm m}}\right)
    \left(\frac{r_{\rm det}}{2~{\rm m}}\right)^2 \left(\frac{131~{\rm m}}{l_{\rm dump}+l_{\rm sh}+l_{\rm dec}}\right)^{2}\notag
\\    
    &\times\left(\frac{|c_{aee}/\Lambda|}{ 0.1~{\rm GeV^{-1}}}\right)^4\left(\frac{m_a}{3\times 10^{-4}~{\rm GeV}}\right)^{12},
\end{align}
where we use some approximations: $\dd {l}_{\gamma}/\dd E_{\gamma}\propto E_{\rm beam}/E_{\gamma}^2$, $E_a\simeq E_{\gamma}$,  $\dd\sigma(\gamma\mathrm{N}\to a\mathrm{N})/\dd\theta_a\propto \left|g_{a\gamma\gamma}\right|^2=\left|c_{aee}/\Lambda\cdot B_1(x_e)\right|^2$, $(l_a^{\rm lab})^{-1}\propto m_a \Gamma(a\to\gamma\gamma)/E_a^{\rm lab}$, $\Gamma(a\to\gamma\gamma)\propto \left|c_{aee}/\Lambda\cdot B_1(x_e)\right|^2m_a^3$, and $B_1(x_e)\simeq -m_a^2/12m_e^2$.
For $ 2m_e \leq m_a\leq \mathcal{O}(10^{-2})~{\rm GeV}$, the decay mode into a electron pair makes dominant in the total decay width of the ALP, and $\theta_1$ is still greater than $\theta_2$ and $\theta_3$.
Similar to the above case, the number of signals is approximately calculated as
\begin{align}
    N_{\rm signal}&\sim \left(\frac{E_{\rm beam}}{125~{\rm GeV}}\right)\left(\frac{N_{e^{\pm}}}{4\times 10^{21}}\right) \left(\frac{l_{\rm dec}}{50~{\rm m}}\right)
    \left(\frac{r_{\rm det}}{2~{\rm m}}\right)^2\left(\frac{131~{\rm m}}{l_{\rm dump}+l_{\rm sh}+l_{\rm dec}}\right)^{2}\notag\\
    &\times\left(\frac{|c_{aee}/\Lambda|}{2\times 10^{-5}~{\rm GeV^{-1}}}\right)^4\left(\frac{m_a}{2\times 10^{-3}~{\rm GeV}}\right)^{2},
\end{align}
where we use following approximations: $\dd{l}_{\gamma}/\dd E_{\gamma}\propto E_{\rm beam}/E_{\gamma}^2$, $E_a\simeq E_{\gamma}$,  $\dd\sigma(\gamma\mathrm{N}\to a\mathrm{N})/\dd\theta_a\propto \left|g_{a\gamma\gamma}\right|^2=\left|c_{aee}/\Lambda\cdot B_1(x_e)\right|^2$, $(l_a^{\rm lab})^{-1}\propto m_a \Gamma(a\to e^+ e^-)/E_a^{\rm lab}$, $\Gamma(a\to e^+ e^-)\propto \left|c_{aee}/\Lambda\right|^2\cdot m_a$, and $B_1(x_e)\simeq 1$.
For $ \mathcal{O}(10^{-2}) \leq m_a\lesssim 1.2\times 10^3 m_e$, the decay mode into a electron pair still makes dominant in the total decay width of the ALP, and $\theta_3$ tends to be greater than $\theta_1$ and $\theta_2$. 
As mentioned in Ref.~\cite{Sakaki:2020mqb}, the minimum value of incident photon energy is proportional to $m_a$ according to the acceptance in Eq.~\eqref{eq:acc}.
Using the same approximations as just above, the number of signals is obtained as
\begin{align}
    N_{\rm signal}&\sim \left(\frac{E_{\rm beam}}{125~{\rm GeV}}\right)\left(\frac{N_{\rm e^{\pm}}}{4\times 10^{21}}\right)\left(\frac{l_{\rm dec}}{50~{\rm m}}\right)
    \left(\frac{r_{\rm det}}{2~{\rm m}}\right)^2 \left(\frac{131~{\rm m}}{l_{\rm dump}+l_{\rm sh}+l_{\rm dec}}\right)^{2}\notag\\
    &\times\left(\frac{|c_{aee}/\Lambda|}{7\times 10^{-6}~{\rm GeV^{-1}}}\right)^4.
\end{align}
If the ALP mass exceeds $1.2\times 10^3 m_e$, the decay mode into photons becomes comparable to that of an electron pair, and the above formula is slightly modified.

In the larger coupling regions, the contour lines are determined by the acceptance in Eq.~\eqref{eq:acc}.
The contour lines are characterized by
\begin{align}
    \left|c_{all}\right|^2\propto
    \begin{cases}
     &m_a^{-2}~~{\rm for}~m_a<1.2\times 10^3 m_e,
     \\
     &m_a^{-4}~~{\rm for}~1.2\times 10^3 m_e\leq m_a,
    \end{cases}\label{eq:upplineALPPr}
\end{align}
where $\Gamma_a\propto  \left|c_{all}\right|^2 m_a$ for $m_a<1.2\times 10^3 m_e$, and $\Gamma_a\propto  \left|c_{all}\right|^2 m_a^3$ for $1.2\times 10^3 m_e\leq m_a$ are used.

\item Bremsstrahlung.
In the small coupling regions, the decay probability in Eq.~\eqref{eq:decayP} behave as $\dd P_{\rm dec}/\dd z\simeq (l^{(\rm lab)}_a)^{-1}$.
For $m_a < 2m_e$, the decay length is dominated by the decay mode into photons, and $\theta_1$ is greater than $\theta_2$ and $\theta_3$.
Then, the number of signals is approximately obtained as
\begin{align}
    N_{\rm signal}&\sim \left(\frac{E_{\rm beam}}{125~{\rm GeV}}\right) \left(\frac{N_{e^{\pm}}}{4\times 10^{21}}\right)\left(\frac{r_{\rm det}}{2~{\rm m}}\right)\left(\frac{l_{\rm dec}}{50~{\rm m}}\right)\left(\frac{131~{\rm m}}{l_{\rm dump}+l_{\rm sh}+l_{\rm dec}}\right)\notag
\\  
&\times\left(\frac{\left|c_{aee}/\Lambda\right|}{5\times 10^{-3}~{\rm GeV}^{-1}}\right)^4 \left(\frac{m_a}{3\times 10^{-4}~{\rm GeV}}\right)^8, 
\end{align}
where $\dd l_{e^{\pm}}/\dd E_{e^{\pm}}\simeq (\dd l_{e^{\pm}}/\dd E_{e^{\pm}})_{\rm shower} \propto E_{\rm beam}/E_{e^{\pm}}^2$, $\sigma(e^{\pm}\mathrm{N}\to e^{\pm} a\mathrm{N})\propto \left|c_{aee}/\Lambda\right|^2$ for $m_a\ll 2 m_e$, $(l^{\rm (lab)}_a)^{-1}\propto \left|c_{aee}/\Lambda\cdot B_1(x_e)\right|^2 m_a^4/E_a^{\rm lab}$, and $B_1(x_e)\simeq -m_a^2/12 m_e^2$ are used. 
As shown in Eqs.~\eqref{eq:Lag-ALP-agg} and \eqref{eq:loop-func}, the axion-photon coupling constant is proportional to $m_a^2$ in the low mass region, for which the lower side of the solid contour lines in Figs.~\ref{fig:contour_ALPscale} and \ref{fig:contour_ALPscale2} do not depend on the ALP mass.
For $2 m_e\leq m_a <\mathcal{O}(10^{-2})~{\rm GeV}$, the decay into an electron pair makes dominant in the total decay width of the ALP, and $\theta_1$ is still greater than $\theta_2$ and $\theta_3$. 
Then, the number of signals behave as
\begin{equation}
\begin{split}
     N_{\rm signal}&\sim
\left(\frac{E_{\rm beam}}{125~{\rm GeV}}\right) \left(\frac{N_{e^{\pm}}}{4\times 10^{21}}\right)\left(\frac{l_{\rm dec}}{50~{\rm m}}\right)\left(\frac{r_{\rm det}}{2~{\rm m}}\right)
\left(\frac{131~{\rm m}}{l_{\rm dump}+l_{\rm sh}+l_{\rm dec}}\right)
\\&\quad
\times\left(\frac{\left|c_{aee}/\Lambda\right|}{10^{-5}~{\rm GeV}^{-1}}\right)^4,
\end{split}\label{eq:appbrem2}
\end{equation}
where $\dd l_{e^{\pm}}/\dd E_{e^{\pm}}\simeq (\dd l_{e^{\pm}}/\dd E_{e^{\pm}})_{\rm shower} \propto E_{\rm beam}/E_{e^{\pm}}^2$, $\sigma(e^{\pm}\mathrm{N}\to e^{\pm} a\mathrm{N})\propto \left|c_{aee}/\Lambda\right|^2 m_a^{-2}$, and $(l^{\rm (lab)}_a)^{-1}\propto \left|c_{aee}/\Lambda\right|^2 m_a^2/E_a^{\rm lab}$ are used.  
Similar to the Primakoff process, in the larger coupling regions, the upper lines of the contour plot behave as Eq.~\eqref{eq:upplineALPPr}. 

\end{enumerate}

\subsection{Light scalar bosons}\label{sec:result-scalar}

The Lagrangian related to the CP-even scalar particle $S$ is written as follows~:
\begin{align}
\label{eq:Lag-scalar}
    \mathcal{L} =
    \frac{1}{2} (\partial_\mu S)^2 - \frac{1}{2} m_S^2 S^2 - \sum_{\ell = e,\mu,\tau} g_\ell S \bar{\ell} \ell-\frac{1}{4}g_{S\gamma\gamma}S F_{\mu\nu}F^{\mu\nu}~
\end{align}
with $m_S$ being the mass of the scalar particle, $g_\ell$ the coupling to the SM charged leptons, and $g_{S\gamma\gamma}$ the coupling to the photons.
Here, we assume that the coupling $g_\ell$ is proportional to the mass of the charged lepton and then satisfies $g_e/m_e = g_\mu/m_\mu = g_\tau/m_\tau$.
Also, we assume that the coupling 
to photons arises by the loop corrections from the electron, i.e.,\cite{Marsicano:2018vin} 
\begin{align}
\label{eq:Lag-scalar-Sgg}
   g_{S\gamma\gamma}=\frac{\alpha}{\pi}\left|\sum_{\ell=e,\mu,\tau}\frac{g_{\ell}}{m_{\ell}} x_{\ell}\cdot B_2(x_{\ell})\right|,
\end{align}
where $x_{\ell}=4 m_{\ell}^2/m_S^2$, and the loop function is $B_2(\tau)=1-(\tau-1)\cdot f(\tau)^2$.

The scalar particles are produced by all the three production mechanisms in Fig.~\ref{fig:diag}; the production cross sections are summarized in Appendix~\ref{sec:xs}. The decay width of the scalar particles is obtained as
\begin{align}
    \Gamma(S \to \ell \bar{\ell}) = \frac{g_\ell^2 m_S}{8 \pi} \left( 1 - \frac{4 m_\ell^2}{m_S^2} \right)^\frac{3}{2}~,
    \\
    \Gamma(S\to\gamma\gamma) =\frac{\alpha^2 m_S^3}{64\pi^3}\left|\sum_{\ell=e,\mu,\tau}\frac{g_{\ell}}{m_{\ell}}x_{\ell}\cdot B_2(x_{\ell})\right|^2.
\end{align}

\begin{figure}[!t]
\centering
\electronpositronfigures{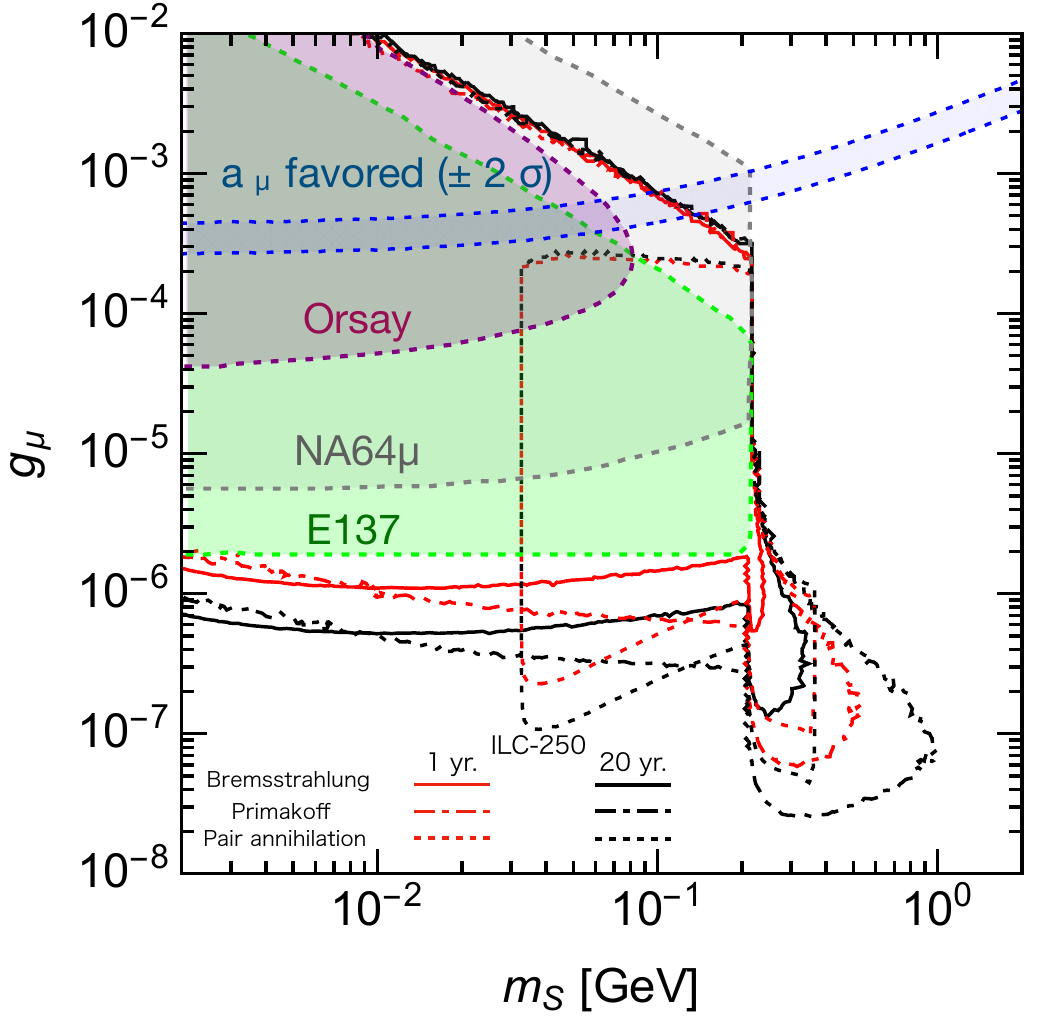}
\caption{%
The red and black curves show the perspectives of sensitivity for ILC-250 at 95\% C.L. with 1- and 20-year statistics.
The solid lines are for the bremsstrahlung process, the dash-dotted lines are for the Primakoff process, and the dotted lines are for the pair annihilation of positrons.
The shaded blue region is a constraint from muon $g-2$ and the shaded gray region is a perspective of a sensitivity of the NA64$\mu$ experiment.   
}
\label{fig:contour_scalar}
\end{figure}

The simulated results are shown in Fig.~\ref{fig:contour_scalar}.
The red and black curves show the expected sensitivity for 95\% C.L. exclusion of ILC-250 with 1- and 20-year statistics.
The solid lines show the expected exclusion region by bremsstrahlung events
\footnote{The sensitivity of the ILC electron beam dump experiment in this process was studied by some of the authors in Ref.~\cite{Sakaki:2020mqb}, where muons as well as electrons are considered as an initial particle in the electromagnetic shower.}.
The dash-dotted (dashed) lines correspond to the exclusion by events originating the Primakoff (pair-annihilation) process.
We show the region excluded by Orsay~\cite{Davier:1989wz} (E137~\cite{Bjorken:1988as}) experiments by purple (green) dashed lines and the expected sensitivity of NA64$\mu$ experiments~\cite{Gninenko:2300189} by gray dashed lines, which are provided in Ref.~\cite{Chen:2017awl}.
We also illustrated the region favored by the muon $g-2$ anomaly by blue dashed lines\footnote{%
  We here considered the SM prediction in Ref.~\cite{Aoyama:2020ynm} together with the measured value~\cite{Bennett:2006fi,Abi:2021gix}.
Note that the lattice calculation of the leading-order hadronic vacuum polarization in Ref.~\cite{Borsanyi:2020mff} suggests the measured value is consistent with SM prediction.
}.

It is found that the ILC beam dump experiment will have higher sensitivity than previous experiments, especially for the small coupling regions.
For the large coupling region, the NA64$\mu$ experiments will be more sensitive, but the region favored by the muon $g-2$ anomaly will be searched for by the bremsstrahlung events~\cite{Sakaki:2020mqb} and the Primakoff events.

We comment on the traditional dark-Higgs scenario, where the dark-Higgs boson as a light scalar has couplings with quarks in addition to leptons.
Below the pion threshold ($m_{\text{dark higgs}} < 2m_\pi$), the contour plot for the dark higgs scenario is almost the same as Fig.~\ref{fig:contour_scalar}.
Above the pion threshold, the decay mode into two pions opens, and the contours would behave just like the region above the muon threshold in Fig.~\ref{fig:contour_scalar}.

Let us explore the results in details, highlighting each of the processes.

\begin{enumerate}[(a)]
\item Pair-annihilation.
Similar to the case of the dark photon and ALP production, with the minimum allowed positron energy $E_{e^+,{\rm min}}\simeq 16~{\rm MeV}\times (l_{\rm dump}+l_{\rm sh}+l_{\rm dec})/r_{\rm det}$, the minimum allowed mass is determined as $m_{S,{\rm min}}\simeq \sqrt{2 m_e E_{e^+,{\rm min}}}$. 
Focusing on the small coupling regions, the number of events in both the positron and electron beam dump experiment for $m_S\leq 2 m_{\mu}$ is approximately calculated as
\begin{align}
    N_{\rm signal}\sim \left(\frac{E_{\rm beam}}{125~{\rm GeV}}\right)\left(\frac{N_{e^{\pm}}}{4\times 10^{21}}\right)\left(\frac{l_{\rm dec}}{50~{\rm m}}\right)\left(\frac{g_{\mu}}{5\times 10^{-7}}\right)^4\left(\frac{0.1~{\rm GeV}}{m_S}\right)^4.
\end{align}
For $2 m_{\mu}\leq m_S$, the decay mode into a muon pair opens, and the possibility of decaying particles passing through the shield increases. Consequently, constraint regions in smaller coupling arise.

In the larger coupling regions, the upper contour lines behave as 
\begin{align}
    \frac{m_S \Gamma_S}{E_S^{\rm lab}}(l_{\rm dump}+l_{\rm sh})\sim {\rm const}.\label{eq:upplineS}
\end{align}
Using $E_S^{\rm lab}\simeq m_S^2/2 m_e$, and $\Gamma_S \propto g_{\mu}^2 m_S$, Eq.~\eqref{eq:upplineS} becomes $g_{\mu}^2\sim {\rm const}$.

\item Primakoff process. 
For $2 m_e\leq m_S <\mathcal{O}(10^{-2})~{\rm GeV}$, the decay probability of $S$ is dominated by the decay mode into an electron pair, and $\theta_1$ is grater than $\theta_2$ and $\theta_3$.
Then, the minimum value of the incident photon is determined as $E_{i,{\rm min}}=8~{\rm MeV}\times \left(l_{\rm dump}+l_{\rm sh}+l_{\rm dec}\right)/r_{\rm det}$ because of the acceptance in Eq.~\eqref{eq:acc}, and the number of signals in the small coupling regions is approximately obtained as
\begin{align}
    N_{\rm signal}&\sim \left(\frac{E_{\rm beam}}{125~{\rm GeV}}\right)\left(\frac{N_{\rm \pm}}{4\times 10^{21}}\right)\left(\frac{r_{\rm det}}{2~{\rm m}}\right)^2 \left(\frac{l_{\rm dec}}{50~{\rm m}}\right)\left(\frac{131~{\rm m}}{l_{\rm dump}+l_{\rm sh}+l_{\rm dec}}\right)^2\notag
\\    
    &\times\left(\frac{g_{\mu}}{2\times 10^{-6}}\right)^4\left(\frac{m_S}{2\times 10^{-3}~{\rm GeV}}\right)^2,
\end{align}
where $\dd P_{\rm dec}/\dd z\simeq 1/l^{({\rm dec})}_S$, $\dd l_{\gamma}/\dd E_{\gamma}\propto E_{\rm beam}/E_{\gamma}^2$, $E_S\simeq E_{\gamma}$, 
$\dd\sigma(\gamma\mathrm{N}\to S\mathrm{N})/\dd\theta_S \propto g_{\mu}^2$, and $(l_S^{({\rm dec})})^{-1}\propto g_{\mu}^2 m_S^2 $.

For $\mathcal{O}(10^{-2})~{\rm GeV}\leq m_S$, $\theta_2$ and $\theta_3$ becomes greater than $\theta_1$, and the minimum value of the incident is proportional to $m_S$.
Then, the number of signals is
\begin{align}
        N_{\rm signal}\sim \left(\frac{E_{\rm beam}}{125\GeV}\right)\left(\frac{N_{\rm EOT}}{4\times 10^{21}}\right)\left(\frac{l_{\rm dec}}{50\,{\rm m}}\right)\left(\frac{r_{\rm det}}{2\,{\rm m}}\right)^2 \left(\frac{131~{\rm m}}{l_{\rm dump}+l_{\rm sh}+l_{\rm dec}}\right)^2\Bigl(\frac{g_{\mu}}{7\times 10^{-7}}\Bigr)^4.
\end{align}

In the larger coupling regions, the upper contour lines are characterized by the acceptance in Eq.~\eqref{eq:acc}.
Using $\Gamma_S\propto g_{\mu}^2$, Eq.~\eqref{eq:acc} behave as $g_{\mu}^2 m_S^2\sim {\rm const.}$.

\item Bremsstrahlung.

For $2 m_e\leq m_S$, the number of signals in the small coupling regions is approximately calculated as
\begin{align}
    N_{\rm signal}\sim \left(\frac{E_{\rm beam}}{125~{\rm GeV}}\right)\left(\frac{N_{e^{\pm}}}{4\times 10^{21}}\right)\left(\frac{l_{\rm dec}}{50~{\rm m}}\right)\left(\frac{r_{\rm det}}{2~{\rm m}}\right)\left(\frac{131~{\rm m}}{l_{\rm dump}+l_{\rm sh}+l_{\rm dec}}\right)\left(\frac{g_{\mu}}{10^{-6}}\right)^4,
\end{align}
where we used $\dd P_{\rm dec}/\dd z\simeq 1/l^{(\rm lab)}_S$, $\dd l_{e^{\pm}}/\dd E_{e^{\pm}}\simeq (\dd l_{e^{\pm}}/\dd E_{e^{\pm}})_{\rm shower}\propto E_{\rm beam}/E_{e^{\pm}}^2$, $\sigma(e^{\pm}\mathrm{N}\to e^{\pm}S\mathrm{N})\propto g_{\mu}^2 m_S^{-2}$, and $(l^{(\rm lab)}_S)^{-1}\propto g_{\mu}^2 m_S^2$.

In the larger coupling regions, the shape of the contour line is determined by the acceptance in Eq.~\eqref{eq:acc}, and the upper contour line is the same as the case of the Primakoff process.  

\end{enumerate}

\section{Summary}
\label{sec:summary}
We performed a feasibility study of the ILC beam dump experiments.
We update the results in Ref.~\cite{Sakaki:2020mqb} with three improvements.
Firstly, we considered both the electron and positron beam dumps and compare their capability.
Three models, i.e., dark photons, ALPs, and light scalar bosons are considered as the BSM scenarios.
Finally, all the relevant processes for the production of the light particles, shown in Fig.~\ref{fig:diag}, are included in the analyses.
We also collected the formulae useful for studies of beam dump experiments in appendices.

The results are collected in Figs.~\ref{fig:contour_DP}--\ref{fig:contour_scalar}.
We found that the positron beam dump experiment is expected to have a slightly higher sensitivity than the electron beam dump if the BSM particles are produced by the pair annihilation process, which is in particular notable in Fig.~\ref{fig:contour_DP}.
For the other scenarios, which are searched for mainly through the bremsstrahlung or Primakoff processes, they have similar sensitivity to the new physics.

In all the scenarios, the ILC beam dump experiment is expected to be sensitive to unexplored parameter regions, even with a 1-year run.
Compared to future experiments, its competence will be comparable to the SHiP experiment for the dark photon model, while for the scalar scenario it will be in particular promising in the smaller coupling region and complementary to the NA64$\mu$ experiment, which is sensitive to the larger coupling region.
We therefore conclude that the ILC beam dump experiment is strongly motivated.

Our estimation is subject to several assumptions and therefore further studies with Monte Carlo simulations are required.
In particular, the angular acceptance is approximated by Eqs.~\eqref{eq:acc} and \eqref{eq:rPerp} but precise evaluation is needed.
We also note that models with large couplings to muons but tiny to electrons are not studied in this work, mainly because we have neglected events originating muons in the electromagnetic showers or with decays of hypothetical particles into muons within the lead shield.

In summary, based on its exploration capability complementary to the ILC $e^+e^-$ collision,
we conclude that the ILC beam dump experiments are necessary to exploit the full ability of the high-energy electron and positron beams, which are not inexpensive, and further dedicated studies with Monte Carlo simulations are highly expected.

\section*{Acknowledgment}
This work was supported by JSPS KAKENHI Grant Number JP19J13812 [KA].
The authors thank the Yukawa Institute for Theoretical Physics at Kyoto University, where this work was initiated during the YITP-W-20-08 on ``Progress in Particle Physics 2020''.

\appendix

\section{Track length in electromagnetic showers}
\label{sec:tracklength}

To describe electromagnetic showers in material, it is useful to utilize the track length $l_i$ of a particle $i$, which is defined by the total flight length of the particles $i$ produced in the shower, including the primary track itself if $i$ is the incident particle.
In Eq.~\eqref{eq:sig}, the track lengths in electromagnetic showers in water induced by $125\GeV$ $e^\pm$ beams are used.

A normalized track length is defined by
\begin{equation}
 \hat l_i = \frac{\rho}{X_0} l_i,
\end{equation}
which is dimensionless; $X_0$ and $\rho$ are the radiation length and the density of the corresponding thick target material; for water beam dumps, $X_0=36.08\unit{g/cm^2}$ and $\rho=1.00\unit{g/cm^3}$.

\begin{figure}[!t]
\begin{center}
\includegraphics[width=9.0cm, bb=0 0 650 480]{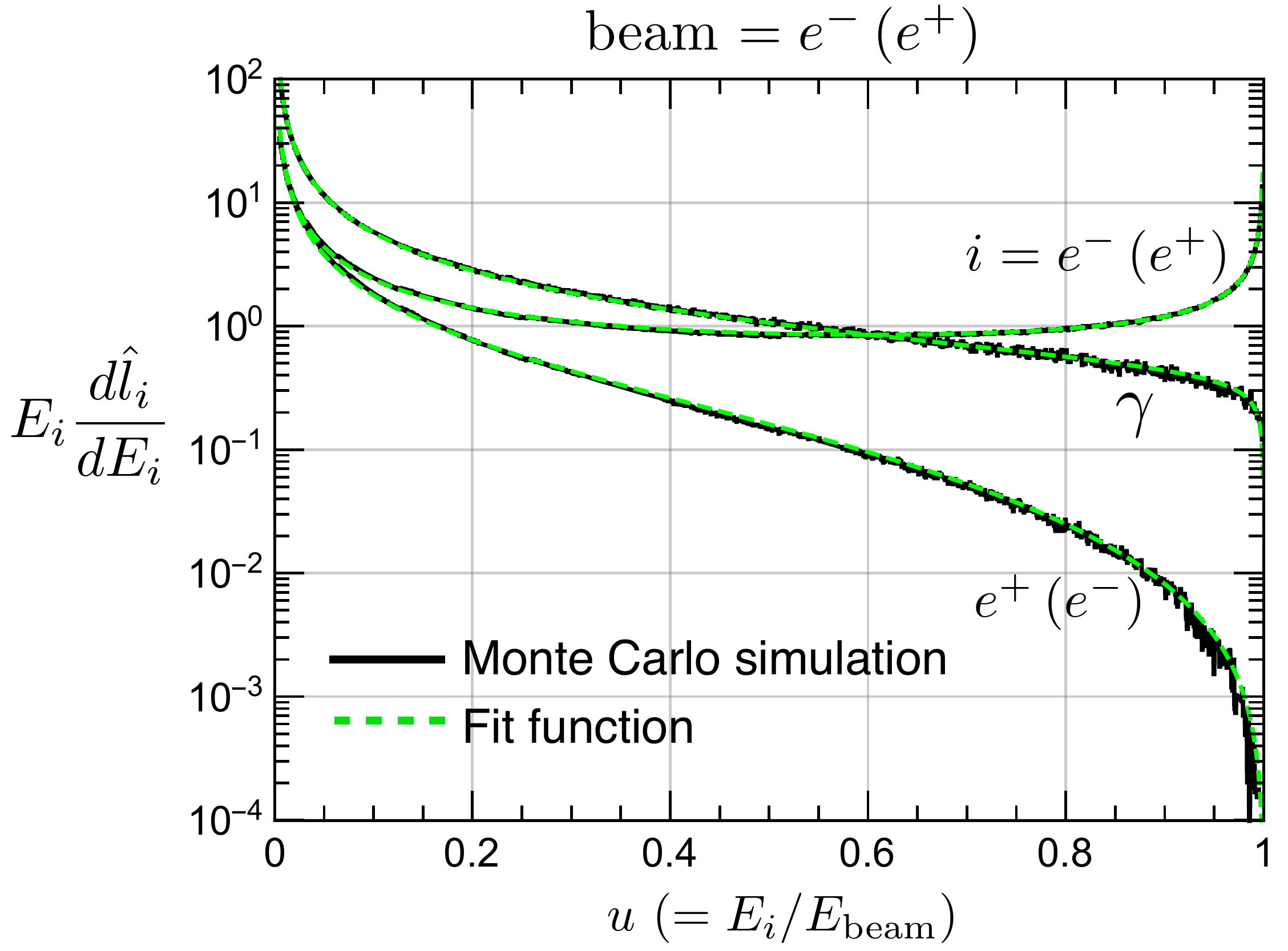}
\caption{Normalized differential track lengths for particle $i$ $(=e^-, e^+, \gamma)$ in a thick target $(\gtrsim 10X_0)$ induced by $e^-$ or $e^+$ beam.}
\label{fig:track_length}
\end{center}
\end{figure}

We estimate the track lengths by Monte Carlo simulations with {\tt EGS5}~\cite{Hirayama:2005zm} code embedded in {\tt PHITS~3.23}~\cite{Sato:2018} for an electron beam of 100\,GeV and an oxygen target of $30X_0$.
The result is shown in Fig.~\ref{fig:track_length} together with the fitting functions.
The results are double-checked by {\tt Geant4}~\cite{Agostinelli:2002hh} simulation for a water target of $11\unit{m}$ length and $125\GeV$ $e^\pm$ beams.

Our fitting functions are given by
\begin{equation}
 \frac{\dd l_{i}}{\dd E_i} = \frac{X_0}{\rho \Ebeam}\frac{\dd \hat l_{i}}{\dd u},
\end{equation}
where $u=E_i/\Ebeam$ with $\Ebeam$ being the energy of the incident particle, and
\begin{alignat}{2}
u\frac{\dd \hat l_{\gamma}}{\dd u}&=
   \frac{0.572}{u}+0.067\ln(1-\sqrt{u})
&&\text{(photon from $e^\pm$ beam)},\\
u\frac{\dd \hat l_{e}}{\dd u}&=
\biggl(u\frac{\dd \hat l_{e}}{\dd u}\biggr)\w{primary}
+
\biggl(u\frac{\dd \hat l_{e}}{\dd u}\biggr)\w{shower}
\quad&&\text{($e^\pm$ from $e^\pm$ beam)},
\\
u\frac{\dd \hat l_{e}}{\dd u}&=
\biggl(u\frac{\dd \hat l_{e}}{\dd u}\biggr)\w{shower}
&&\text{($e^\mp$ from $e^\pm$ beam)},
\end{alignat}
with
\begin{align}
\biggl(u\frac{\dd \hat l_{e}}{\dd u}\biggr)\w{primary}
 &= 0.581 + 0.131 \left( \frac{u}{1-u} \right)^{0.7},\\
\biggl(u\frac{\dd \hat l_{e}}{\dd u}\biggr)\w{shower}&=
  \frac{1-u}{u}(0.199-0.155u^2).
\end{align}
These formulae work for targets with enough thickness $(\gtrsim 10X_0)$, because the target thickness dependency of track length is small in that case.
These are also valid for the beam energy at which the electromagnetic shower develops sufficiently ($>\mathcal{O}(1)$ GeV) and for different materials by using the corresponding radiation length.\footnote{%
  We explicitly checked that these fitting functions are applicable for $500\GeV$ $e^\pm$ beams.
}

\section{Production cross sections of the new light particles}
\label{sec:xs}
In this work, we consider the production of hypothetical new particles $X$ through pair-annihilation, Primakoff, and bremsstrahlung processes (cf.\ Fig.~\ref{fig:diag}).
We here collect the production cross sections used in our analysis; together with Eqs.~\eqref{eq:pa}--\eqref{eq:br}, one can calculate the expected number of events.

The following formulae are given in the lab frame, where an incoming particle $i$ hits a target particle $j$ being at rest; $E_i$ denotes the lab-frame energy of the incoming particle.
Dark photons, ALPs, and new scalars are denoted by $A'$, $a$, and $S$, respectively, and their masses and decay widths are denoted by $m_X$ and $\Gamma_X$.
The model parameters are found in Eqs.~\eqref{eq:Lag-DP}, \eqref{eq:Lag-ALP}--\eqref{eq:Lag-ALP-agg}, and \eqref{eq:Lag-scalar}--\eqref{eq:Lag-scalar-Sgg}.
For Primakoff and bremsstrahlung processes, a target nucleus is denoted by $\mathrm N$ with the atomic number $Z$ and mass number $A$.

\subsection{Pair annihilation production}

The cross sections of the resonant annihilation process (Fig.~\ref{fig:diag-pa}) are given by,
respectively for the dark photon, ALP, and light scalar,~\cite{Marsicano:2018krp,Bauer:2018uxu}
\begin{align}
\label{eq:sigma_ann_DP}
    \sigma(e^+e^-\to A')
&= \frac{12\pi}{m_{A'}^2} \frac{\Gamma_{A'}^2/4}{(\sqrt{s}-m_{A'})^2 + \Gamma_{A'}^2/4}~, \\
\label{eq:sigma_ann_ALP} 
    \sigma(e^+e^-\to a)
&= \frac{\sqrt{s} m_e^2}{8\pi} \left(\frac{c_{aee}}{\Lambda}\right)^2 \frac{4 \pi \Gamma_a}{(s-m_a^2)^2 + m_a^2 \Gamma_a^2}~, \\
\label{eq:sigma_ann_scalar}
    \sigma(e^+e^-\to S)
&= \frac{\sqrt{s}m_e^2}{8\pi} \frac{g_e^2}{m_e^2} \frac{4 \pi \Gamma_S}{(s-m_S^2)^2 + m_S^2 \Gamma_S^2}~,
\end{align}
where $s$ is the center-of-mass energy squared.
In this work, we assume that the total decay widths of the new particles are small enough to pass through the lead shield.
We thus use the narrow-width approximation to obtain
\begin{align}
\label{eq:sigma_ann_DP_limit}
    \sigma(e^+e^-\to A')
 &\simeq \frac{2\pi^2 \alpha \epsilon^2}{m_e} ~\ddelta \left( E_i - \frac{m_{A'}^2}{2m_e} + m_e \right)~, \\
\label{eq:sigma_ann_ALP_limit}
    \sigma(e^+e^-\to a)
 &\simeq \frac{\pi m_e}{4} \left(\frac{c_{aee}}{\Lambda}\right)^2 ~\ddelta \left( E_i - \frac{m_a^2}{2m_e} + m_e \right)~, \\
\label{eq:sigma_ann_scalar_limit}
    \sigma(e^+e^-\to S)
 &\simeq \frac{\pi g_e^2}{4 m_e} ~\ddelta \left( E_i - \frac{m_S^2}{2m_e} + m_e \right)~
\end{align}
with $\ddelta(x)$ being the Dirac delta function.

\subsection{Primakoff production}
We used the differential cross sections of the Primakoff production (Fig.~\ref{fig:diag-pr}) calculated with the improved Weizs\"acker-Williams approximation\cite{vonWeizsacker:1934nji,Williams:1935dka,Kim:1973he}.
They are given by~\cite{Tsai:1986tx,Bjorken:1988as,Dusaev:2020gxi,Dobrich:2015jyk}
\begin{align}
    \frac{\dd\sigma\left(\gamma\mathrm{N}\to a\mathrm{N}\right)}{\dd\theta_a}
\simeq \frac{\alpha g_{a\gamma\gamma}^2E_i^4 \theta_a^3}{4 t^2} G_2(t),\label{eq:praeq}
\\
 \frac{\dd\sigma\left(\gamma\mathrm{N}\to S\mathrm{N}\right)}{\dd\theta_S}
\simeq \frac{\alpha g_{S\gamma\gamma}^2E_i^4 \theta_S^3}{4 t^2} G_2(t),\label{eq:prSeq}
\end{align}
where $g_{X\gamma\gamma}$ is the $X$-photon-photon coupling and $\theta_X$ is the angle of the outgoing $X$ with respect to the incoming photon.
The electric form factor squared is given by (cf.\ Ref.~\cite{Bjorken:2009mm})
\begin{equation}
    G_2(t)\approx G_{2,\,\textrm{elastic}}(t) = \left[\left(\frac{a^2 t}{1+a^2 t}\right)\left(\frac{1}{1+t/d}\right)Z\right]^2\,,
\end{equation}
where $a=111Z^{1/3}/m_e$, $d=0.164\GeV^2/A^{2/3}$, and 
\begin{equation}
t=-q^2\simeq E_i^2 \theta_X^2+\frac{m_X^4}{4E_i^2}
\end{equation}
with $q$ denoting the momentum transfer. The outgoing particle $X$ has the energy
\begin{equation}
    E_X\simeq E_i -\frac{E_i^2 \theta_X^2}{2M\w N} -\frac{m_X^4}{8M\w N E_i^2},
\end{equation}
where $M\w N$ is the mass of the target nucleus.
Note that these formulae are derived with the assumptions $m_X \ll M\w N$ and $t \ll M\w N^2$.

\subsection{Bremsstrahlung production}
The differential cross sections of the bremsstrahlung process (Fig.~\ref{fig:diag-br}) calculated with the improved Weizs\"acker-Williams approximation\cite{vonWeizsacker:1934nji,Williams:1935dka,Kim:1973he} are given by~\cite{Liu:2017htz,Bjorken:2009mm,Liu:2016mqv}
\begin{align}
\frac{\dd^2\sigma(e^\pm\mathrm{N}\to e^\pm X\mathrm{N})}{\dd x\,\dd \theta_X} =
    \frac{g_{Xee}^2 \alpha^2}{2 \pi} x (1-x) E_i^2 \beta_X \frac{\mathcal{A}^X|_{t = t\w{min}}}{\tilde{u}^2} \chi\, ,\label{eq:sigma_brems}
\end{align}
where $x=E_X/E_i$ and $\beta_X=\sqrt{1-m_X^2/E_i^2}$.
The coupling $g_{Xee}$ is given by
\begin{equation}
 g_{Xee} =
\begin{cases}
\epsilon e & \text{(dark photon)},\\
-m_e c_{aee}/{\Lambda} & \text{(ALP)},\\
g_e & \text{(scalar)}
\end{cases}
\end{equation}
for each model, and
\begin{equation}
    \tilde{u} = - x E_i^2 \theta_X^2 - m_X^2 \frac{1-x}{x} - m_e^2 x.\\
\end{equation}
The effective flux of photons, $\chi$, is given by
\begin{equation}
    \chi = \int_{t\w{min}}^{t\w{max}} \dd t~ \frac{t - t\w{min}}{t^2} G_2(t)~,
\end{equation}
where $t\w{min} = (m_X^2 / 2 E_i)^2$ and $t\w{max} = m_X^2$.
The amplitude under Weizs\"acker-Williams approximation, $\mathcal{A}^X$ evaluated at $t = t\w{min}$, is given by
\begin{align}
    \mathcal{A}^{A'}|_{t = t\w{min}} &= 2 \frac{2 - 2 x + x^2}{1-x} + 4 (m_{A'}^2 + 2 m_e^2) \frac{\tilde{u} x + m_{A'}^2 (1-x) + m_e^2 x^2}{\tilde{u}^2}~, \\
    \mathcal{A}^a|_{t = t\w{min}} &= \frac{x^2}{1-x} + 2 m_a^2 \frac{\tilde{u} x + m_a^2 (1-x) + m_e^2 x^2}{\tilde{u}^2}~, \\
    \mathcal{A}^S|_{t = t\w{min}} &= \frac{x^2}{1-x} + 2 (m_S^2 - 4 m_e^2) \frac{\tilde{u} x + m_S^2 (1-x) + m_e^2 x^2}{\tilde{u}^2}~,
\end{align}
for the dark photon, ALP, and scalar particle, respectively.

\section{Note on angular distribution of shower particles and the acceptance}
\label{sec:theta-dist}

\begin{figure}[!t]
\centering
\electronpositronfigures{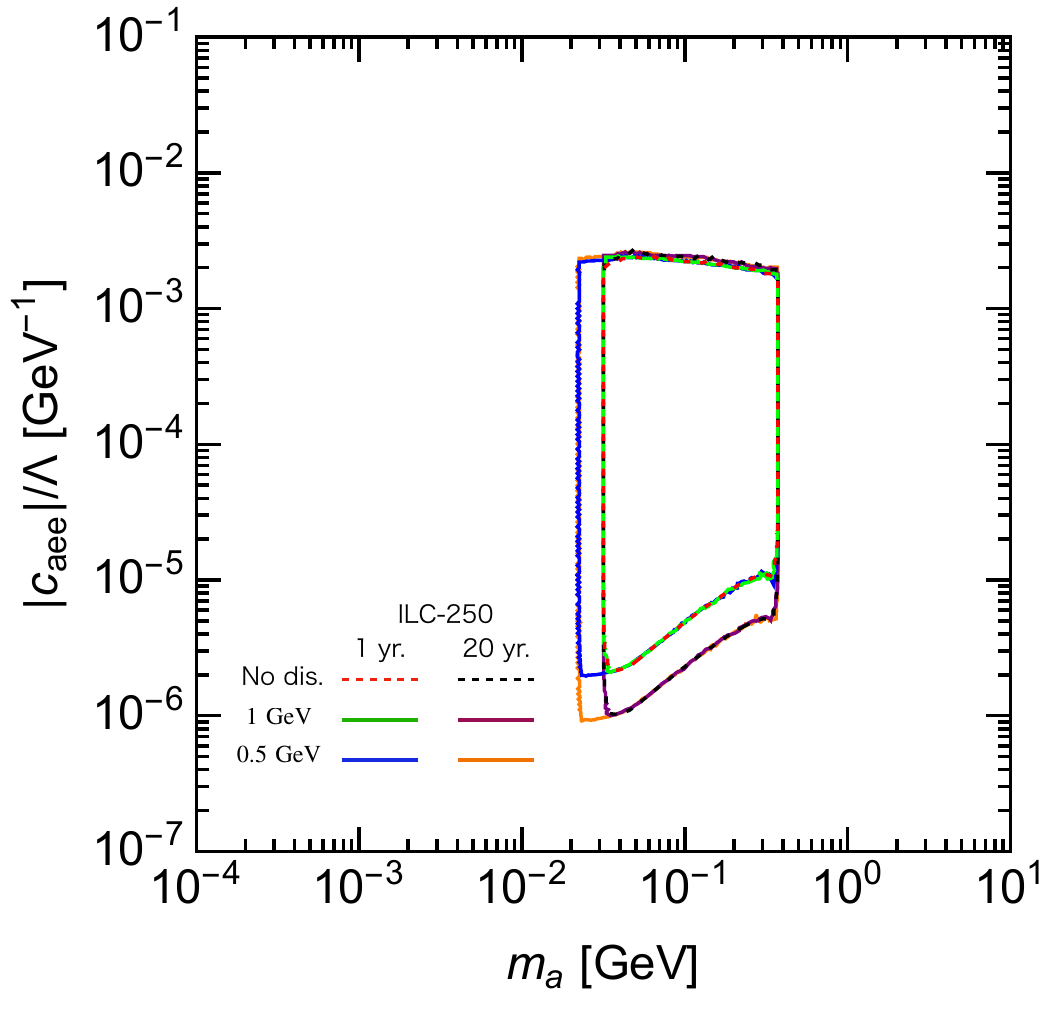}
\caption{%
The red and black dotted lines show the bounds of sensitivity ILC-250 at 95\% C.L. with 1- and 20-year statistics without the effect of the angular distribution of incident positrons.
The green and purple solid lines are the same bounds as the red and black dotted lines but the angular distribution of incident positrons and $E_{\rm th}=1~{\rm GeV}$ are included.
Note that the red and black dotted lines almost overlap the green and purple lines and cannot be distinguished. 
The blue and orange solid lines are the same bounds as the green and purple solid lines but for $E_{\rm th}=0.5~{\rm GeV}$.
}
\label{fig:contour_ALPth}
\end{figure}

In our analyses in Sec.~\ref{sec:results}, we evaluated the angle $\theta_1$ of shower particles by its mean value, Eq.~\eqref{eq:theta1mean}.
As a consequence of this treatment and Eq.~\eqref{eq:acc}, shower photons (electrons or positrons) with the energy $E_{\gamma}<0.52\GeV$ ($E_{e^\pm}<1.05\GeV$) always result in $r_\perp>r\w{det}$ and never contribute to the number of signal events.
For example, the low-mass boundaries of the expected sensitivity for the pair-annihilation processes are determined by this energy threshold.

In reality, $\theta_1$ has a distribution, and shower particles with smaller momentum may pass the angular cut.
Here, we have to take into account that the detector will have a minimal energy for detection, $E\w{th}$, which we did not include in Sec.~\ref{sec:results}.

In Fig.~\ref{fig:contour_ALPth}, we show the effects on the exclusion sensitivity from these different treatments of $\theta_1$.
The result for the ALP model (Case I) is displayed to compare with Fig.~\ref{fig:contour_ALP}, but similar results are obtained for the other models.
For brevity, instead of requiring the resulting SM particles to have $E>E\w{th}$, we require $E_X>E\w{th}$.
The lines show the sensitivity of ILC-250 at 95\% C.L. with 1- and 20-year statistics as in Fig.~\ref{fig:contour_ALP}, but only for the pair-annihilation.
The red and black dotted lines are obtained with Eq.~\eqref{eq:theta1mean} (without $E\w{th}$-requirement) and thus equal to those in Fig.~\ref{fig:contour_ALP}.
The green and purple solid lines are calculated with the $\theta_1$ distribution and requiring $E_X>1\GeV$; they perfectly overlap with the red and black dotted lines as expected.
The blue and orange solid lines are calculated with the $\theta_1$ distribution and requiring $E_X>0.5\GeV$.
It is shown that the low-mass boundaries are sensitive to the energy threshold but the other boundaries are independent of the treatment of $\theta_1$; in particular, the small-coupling boundaries are controlled by the integrated luminosity.

\bibliography{ref}

\end{document}